\documentclass[a4paper,11pt]{amsart}
\usepackage{amsmath,amssymb,amsfonts,latexsym}
\usepackage[english]{babel}
\usepackage[dvipsnames]{xcolor}
\usepackage{multirow}
\usepackage[left=2.5cm,right=2.5cm,top=3.cm,bottom=3.cm]{geometry}
\usepackage{graphicx}
\usepackage{tikz}
\usepackage{pstricks}
\usepackage{color}
\usepackage{pgfplots}
\usepackage{enumitem}
\usepackage{textcomp}
\usepackage{colortbl}
\usepackage{slashbox}
\usepackage{setspace}
\usetikzlibrary{pgfplots.statistics}

\usepackage{arydshln}
\usetikzlibrary{matrix,shapes,arrows,positioning}
\definecolor{myblue}{RGB}{209,221,243}
\definecolor{myblu}{RGB}{110,148,216}
\definecolor{myre}{RGB}{255,188,188}
\definecolor{li}{gray}{0.97}
\definecolor{lt}{gray}{0.7}
\usepackage{tabu}
\usepackage{pifont}
\usepackage{hhline}
\usepackage{algorithm}
\usepackage[noend]{algpseudocode}
\makeatletter
\def\BState{\State\hskip-\ALG@thistlm}
\usetikzlibrary{shadows}
\usepackage{booktabs}

\renewcommand{\algorithmicrequire}{\textbf{Input: }}
\renewcommand{\algorithmicensure}{\textbf{Output: }}
\usetikzlibrary{patterns}
\algrenewcommand\algorithmicindent{1.4em}%
\usetikzlibrary{calc}

\begin{document}
\title[Cuckoo Search Algorithm for the H-P Protein Folding Problem ]{\textbf{\vspace{-3mm} \\}PROTEIN FOLDING SIMULATIONS IN THE HYDROPHOBIC-POLAR MODEL USING A HYBRID CUCKOO SEARCH ALGORITHM}

\author[]{N\MakeLowercase{abil} B\MakeLowercase{oumedine}$^1$,  S\MakeLowercase{adek} B\MakeLowercase{OUROUBI}$^2$\vspace{2mm} \\ \MakeLowercase{\texttt{nboumdine@usthb.dz}$^1$, \texttt{sbouroubi@usthb.dz}$^2$}  \vspace{3mm}\\$^{1,2\,}$ University of Sciences and Technology Houari Boumediene (USTHB),\vspace{1mm}\\
L'IFORCE L\MakeLowercase{aboratory}, \vspace{1mm}\\
 F\MakeLowercase{aculty of }M\MakeLowercase{athematics}, D\MakeLowercase{epartment of} O\MakeLowercase{perations }R\MakeLowercase{esearch},\vspace{1mm} \\ P.B. 32 E\MakeLowercase{l-Alia}, 16111, B\MakeLowercase{ab }E\MakeLowercase{zzouar}, A\MakeLowercase{lgiers}, A\MakeLowercase{lgeria.}
}


\begin{abstract}
    A protein is a linear chain containing a set of amino acids, which folds on itself to create a specific native structure, also called the minimum energy conformation. It is the native structure that determines the functionality of each protein. The protein folding problem (PFP) remains one of the more difficult problems in computational and chemical biology. The principal challenge of PFP is to predict the optimal conformation of a given protein by considering only its amino acid sequence. As the conformational space contains a very large number of conformations, even when addressing short sequences, different simplified models have been developed and applied to make the PFP less complex. In the last few years, many computational approaches have been proposed to solve the PFP. They are based on simplified lattice models such as the hydrophobic-polar model. In this paper, we present a new Hybrid Cuckoo Search Algorithm (HCSA) to solve the 3D-HP protein folding optimization problem. Our proposed algorithm consists of combining the Cuckoo Search Algorithm (CSA) with the Hill Climbing (HC) algorithm. Simulation results on different benchmark sequences are presented and compared to the state-of-the-art algorithms.\vspace{2mm}

\vspace{-3em}
\end{abstract}


\maketitle

\section {Introduction}
 Proteins are very complex molecules that play a vital role in biological processes. The molecular function of any given protein is determined by its specific native three-dimensional structure, also known as  conformation \cite{valastyan_mechanisms_2014, hagerman_sequence_1996}. According to Anfinsen's experiments, proteins fold into their native structure in only a few milliseconds, with minimal free energy, based solely on the information provided by their specific amino acid sequence (i.e., the primary structure) \cite{anfinsen_principles_1973, anfinsen_kinetics_1961}.  However, protein misfolding and aggregation are the cause of several diseases, including Alzheimer's disease, Parkinson's disease, prions, and mad cow disease. \cite{hardy_alzheimers_1992, singleton_-synuclein_2003,   nunnally_prions_2003,   lauren_cellular_2009}. The latter is one of the main motivations for studying the protein folding process. Nevertheless, some experimental methods namely X-ray Crystallography (XC) \cite{bragg_development_1975}, and Nuclear Magnetic Resonance (NMR) \cite{baldwin_crystal_1991}, have been experimented with great success and have been used to predict the native structure of several real proteins. At the same time, these methods are very costly in terms of both equipment and time. Therefore, a variety of computational approaches have been used to address the protein folding problem (PFP).\\
 The question of determining the way in which the amino acid sequence of proteins precisely adopts their native three-dimensional structure is one of the most difficult problems in the fields of biochemistry, computational biology, and computer science, and has been the subject of much work based on simplified models. The challenge is to identify the conformation with the lowest free energy of a given protein using its amino acid sequence information. However, the combinatorial explosion of possible conformations induces a high computational cost, even for very small sequences. Some simplified models have been proposed to reduce the complexity of this problem. The Hydrophobic-Polar (HP) model, proposed by Dill in \cite{dill_theory_1985}, is one of the most frequently used models. This model is characterized by the fact that the folding process is thermodynamically driven by hydrophobic interactions between amino acids, which are key to the development of the native state in proteins \cite{lau_lattice_1989}.
 Based on the simplified H-P lattice model \cite{lau_lattice_1989}, the free energy values are calculated propositionally as a function of the number of hydrophobic contacts, which must be maximized to obtain the optimal native structure. However, even when using the simplified H-P lattice model, the PFP problem is shown to be NP-hard in the two-dimensional Hydrophobic-Polar (2D-HP) model \cite{crescenzi_complexity_1998}, and the three-dimensional Hydrophobic-Polar (3D-HP) model \cite{berger_protein_1998}. In the case where each amino acid has four neighbors in the conformational space, a protein sequence of 100 amino acids has $4^{100}$ possible conformations.  Given the complexity of the PFP problem and the limitation of the experimental methods, a variety of well-known heuristic optimization approaches have been proposed in the literature to solve the PFP problem in  different types of lattices in the 2D-HP and 3D-HP model that can predict well-approximated conformations for the real proteins.
In this work, we introduce an efficient Hybrid Cuckoo Search Algorithm (HCSA) that combines the Cuckoo Search Algorithm (CSA) with a Hill-Climbing algorithm for the H-P protein folding problem in a 3D cubic lattice model.\\
 In the following section, we report and discuss some related work as well as well-known algorithms and heuristics that are used to solve the PFP in both the two-dimensional square (2D-HP) and three-dimensional cubic (3D-HP) lattice models. However, in Section 3, we concentrate on the characterization of the simplified HP lattice model. Further, we define the fitness function (i.e., the energy function) and the search space for this model. In Section 4, we briefly describe a  mathematical formulation in the form of a 0-1 mathematical quadratic  program for the PFP problem in the 3D cubic lattice model. In Section 5 we present and describe our HCSA algorithm in more detail. Furthermore, in Section 6, we summarize and discuss the obtained results  to evaluate the performance and the effectiveness of the suggested algorithm against the state-of-the-art algorithms through two data benchmarks. Finally, in the last section, we present our conclusion and future directions.
 \section{Related Works}
 Actually, there is no exact algorithm for PFP  that finds the optimal conformation for any given protein amino acid sequence. On the other hand, since the presence of the H-P model, various heuristic and metaheuristic techniques in literature have been used to solve the  PFP problem in the 2D-HP square and  3D-HP cubic lattices model. In the 2D-HP model, Unger and Moult \cite{unger_genetic_1993}, suggested the first Genetic Algorithm (GA), using a number of genetic operators to explore the search space. The mutation operator, which was used to change the solutions of the current population, the Monte Carlo steps as acceptance criteria, and the crossover to create new solutions by swapping parts information between pairs of selected solutions. In the proposed GA, the probability of choosing a given solution is directly proportional to its fitness. Later more versions of the GA were proposed in
  \cite{dandekar_folding_1994, krasnogor_protein_1999, konig_improving_1999, lin_efficient_2009}.  In \cite{shmygelska_ant_2002}, the authors have developed the first Ant Colony Optimization (ACO) algorithm for the PFP problem in the 2D-HP square lattice model, which has been followed by an improved version in \cite{shmygelska_improved_2003}. The same approach was also successfully applied to the HP model in the 2D square lattice and the 3D cubic lattice \cite{shmygelska_ant_2005}. The PSO (Particle Swarm Optimization) algorithm has been adopted and used to solve the PFP problem in a 2D square lattice in  \cite{bautu_protein_2010}. Similarly, the application of the ABC (Artificial Bee Colony) and the Honey-Bees Mating Optimization (HBMO) algorithms have been suggested in \cite{lin_using_2012} and \cite{boukra_protein_2012} respectively. In addition, there are several hybrid approaches that have been developed an applied to the PFP in different lattice models, such as the GTS algorithm for the 2D H-P model \cite{jiang_protein_2003}, which is to combine the simple Genetic Algorithm (GA) with Tabu Search (TS). Likewise, the Evolutionary Algorithm (EA) was hybridized with the backtracking algorithm  to solve the PFP in the 3D-HP model \cite{cotta_protein_2003}. Some other metaheuristics that were developed and applied such as, the Memetic Algorithms (MA) \cite{krasnogor_multimeme_2002,islam_clustered_2013}, Immune Algorithm (IA) \cite{cutello_immune_2007, cutello_immune_2005}. In addition, a number of approximation algorithms have been offered for this problem in different lattices. In \cite{hart_fast_1996}, Hart and Istrail presented an algorithm with an approximation ratio of $\frac{1}{4}$ $(25\%)$  for the 2D square lattice and an approximation ratio of $\frac{3}{8}$ $(38\%)$ for the 3D cubic lattice. Later, those approximation ratio were improved to  $\frac{1}{3}$ $(33\%)$ for the 2D square lattice by Newman in \cite{newman_new_2002}, and to $\frac{6}{11}$ $(54\%)$s  for the 3D cubic lattice in \cite{decatur_protein_1996}.
\section{HP Folding in Lattice Model}
\begin{figure}[h]
\definecolor{qqqqff}{rgb}{0.,0.,1.}
\definecolor{ffqqqq}{rgb}{1.,0.,0.}
\definecolor{qqffqq}{rgb}{0.,1.,0.}
\definecolor{wqwqwq}{rgb}{0.3764705882352941,0.3764705882352941,0.3764705882352941}
\begin{tikzpicture}[line cap=round,line join=round,>=triangle 45,x=1.0cm,y=1.0cm]
\clip(40.51814533588907,7.735790042257493) rectangle (48.333543766866775,14.332595614432343);
\draw (44.54887556116093,9.844893067109133) node[anchor=north west] {\footnotesize{H}};
\draw (44.38483421478359,11.403285857693955) node[anchor=north west] {\footnotesize{H}};
\draw (45.22847542472421,12.223492589580705) node[anchor=north west] {\footnotesize{H}};
\draw (43.15452411695352,11.45015481380177) node[anchor=north west] {\footnotesize{H}};
\draw (43.16624135598047,12.821071779955336) node[anchor=north west] {\footnotesize{H}};
\draw (46.63454410795858,12.000865048068587) node[anchor=north west] {\footnotesize{H}};
\draw (45.19332370764335,10.758837711211509) node[anchor=north west] {\footnotesize{H}};
\draw (47.37273016665662,10.501058452618532) node[anchor=north west] {\footnotesize{H}};
\draw (45.99,10.020651652513436) node[anchor=north west] {\footnotesize{H}};
\draw (46.599392390877725,10.641665320941975) node[anchor=north west] {\footnotesize{H}};
\draw (46.55252343476991,9.270748354788408) node[anchor=north west] {\footnotesize{P}};
\draw (45.99,11.391568618667002) node[anchor=north west] {\footnotesize{P}};
\draw (45.99,8.860644988845035) node[anchor=north west] {\footnotesize{P}};
\draw (47.39616464471053,11.918844374879912) node[anchor=north west] {\footnotesize{P}};
\draw (45.25190990277812,13.652995750869037) node[anchor=north west] {\footnotesize{P}};
\draw (43.306848224303906,9.938630979324762) node[anchor=north west] {\footnotesize{P}};
\draw (41.98280021425821,11.579044443098258) node[anchor=north west] {\footnotesize{P}};
\draw (47.37273016665662,9.587113808516154) node[anchor=north west] {\footnotesize{P}};
\draw (41.99451745328516,12.90309245314401) node[anchor=north west] {\footnotesize{P}};
\draw (45.14645475153554,9.411355223111851) node[anchor=north west] {\footnotesize{P}};
\draw (43.69351711219336,8.368520949712986) node[anchor=north west] {\footnotesize{axe x}};
\draw (46.89232336655155,8.544279535117289) node[anchor=north west] {\footnotesize{axe y}};
\draw (40.89309698475157,10.618230842888067) node[anchor=north west] {\footnotesize{axe z}};
\draw (45.263627141805074,12.668747672604939) node[anchor=north west] {\footnotesize{Start}};
\draw (44.31453078062187,10.48) node[anchor=north west] {\footnotesize{End}};
\draw [line width=0.4pt,dash pattern=on 1pt off 1pt,color=wqwqwq] (42.,13.)-- (44.01631540372523,13.850171478897579);
\draw [line width=0.4pt,dash pattern=on 1pt off 1pt,color=wqwqwq] (47.83063248184536,13.415349172622593)-- (47.852470594694445,12.001661464596303);
\draw [line width=0.4pt,dash pattern=on 1pt off 1pt,color=wqwqwq] (42.0684231305906,12.35943759639655)-- (43.91693314489183,13.134619215297086);
\draw [line width=0.4pt,dash pattern=on 1pt off 1pt,color=wqwqwq] (43.91693314489183,13.134619215297086)-- (47.84171713785071,12.697785248910682);
\draw [line width=0.4pt,dash pattern=on 1pt off 1pt,color=wqwqwq] (42.00099901480501,11.65020176655288)-- (43.93680959665851,12.399190499929912);
\draw [line width=0.4pt,dash pattern=on 1pt off 1pt,color=wqwqwq] (43.93680959665851,12.399190499929912)-- (47.852470594694445,12.001661464596303);
\draw [line width=0.4pt,dash pattern=on 1pt off 1pt,color=wqwqwq] (42.,11.)-- (43.93680959665851,11.663761784562737);
\draw [line width=0.4pt,dash pattern=on 1pt off 1pt,color=wqwqwq] (43.93680959665851,11.663761784562737)-- (47.77296478762773,11.266232749229129);
\draw [line width=0.4pt,dash pattern=on 1pt off 1pt,color=wqwqwq] (42.048546678823925,10.23265725736175)-- (44.,11.);
\draw [line width=0.4pt,dash pattern=on 1pt off 1pt,color=wqwqwq] (44.,11.)-- (47.84206589952466,10.583917241350086);
\draw [line width=0.4pt,dash pattern=on 1pt off 1pt,color=wqwqwq] (42.0684231305906,9.536981445527937)-- (43.41491911559936,10.075034968866147);
\draw [line width=0.4pt,dash pattern=on 1pt off 1pt,color=wqwqwq] (42.048546678823925,8.821429181927442)-- (43.91693314489183,9.616487252594657);
\draw [line width=0.4pt,dash pattern=on 1pt off 1pt,color=wqwqwq] (43.380268947191475,8.662417567794)-- (45.188280748486896,9.46749440945095);
\draw [line width=0.4pt,dash pattern=on 1pt off 1pt,color=wqwqwq] (44.01631540372523,8.543158857193918)-- (45.88470186979314,9.338216927861133);
\draw [line width=0.4pt,dash pattern=on 1pt off 1pt,color=wqwqwq] (44.69211476379235,8.483529501893877)-- (46.56756201851053,9.303294258257653);
\draw [line width=0.4pt,dash pattern=on 1pt off 1pt,color=wqwqwq] (45.28840831679274,8.423900146593835)-- (47.17667123462733,9.23883466902773);
\draw [line width=0.4pt,dash pattern=on 1pt off 1pt,color=wqwqwq] (43.91693314489183,9.616487252594657)-- (47.84206589952466,9.154024335031758);
\draw [line width=0.4pt,dash pattern=on 1pt off 1pt,color=wqwqwq] (47.792841239394406,9.894757577328184)-- (43.93208844003778,10.331718385080281);
\draw [line width=2.pt,color=qqqqff] (45.24725627003013,12.266150228521623)-- (45.26552966374888,13.704257400946888);
\draw [line width=2.pt,color=qqqqff] (45.26552966374888,13.704257400946888)-- (43.38255561863366,12.91052541396301);
\draw [line width=2.pt,color=qqqqff] (43.38255561863366,12.91052541396301)-- (42.,13.);
\draw [line width=2.pt,color=qqqqff] (42.,13.)-- (42.00099901480501,11.65020176655288);
\draw [line width=2.pt,color=qqqqff] (42.00099901480501,11.65020176655288)-- (45.99749611216069,11.450219140469242);
\draw [line width=2.pt,color=qqqqff] (45.99749611216069,11.450219140469242)-- (45.996444258330776,8.339859280010607);
\draw [line width=2.pt,color=qqqqff] (47.84206589952466,9.154024335031758)-- (47.852470594694445,12.001661464596303);
\draw [line width=2.pt,color=qqqqff] (47.852470594694445,12.001661464596303)-- (46.629706887503126,12.125799912026897);
\draw [line width=2.pt,color=qqqqff] (46.629706887503126,12.125799912026897)-- (46.56756201851053,9.303294258257653);
\draw [line width=2.pt,color=qqqqff] (46.56756201851053,9.303294258257653)-- (45.188280748486896,9.46749440945095);
\draw [line width=2.pt,color=qqqqff] (45.188280748486896,9.46749440945095)-- (45.23502077410504,10.866251421990667);
\draw [line width=2.pt,color=qqqqff] (45.23502077410504,10.866251421990667)-- (43.41491911559936,10.075034968866147);
\draw [line width=2.pt,color=qqqqff] (43.41491911559936,10.075034968866147)-- (44.61358021931038,9.94367484791151);
\draw [line width=0.4pt,color=wqwqwq] (42.00099901480501,11.65020176655288)-- (42.00099901480501,8.832907043103564);
\draw [line width=0.4pt,color=wqwqwq] (45.996444258330776,8.339859280010607)-- (42.00099901480501,8.832907043103564);
\draw [line width=0.4pt,dash pattern=on 1pt off 1pt,color=wqwqwq] (44.006555031648354,13.84605605630655)-- (47.83067588908211,13.412539210175437);
\draw [line width=0.4pt,dash pattern=on 1pt off 1pt,color=wqwqwq] (46.00043486637473,13.620022275016733)-- (46.000684448716264,9.370977669178723);
\draw [line width=0.4pt,dash pattern=on 1pt off 1pt,color=wqwqwq] (44.60597335925089,13.778103720963827)-- (44.61260895672563,9.497228541994575);
\draw [line width=0.4pt,dash pattern=on 1pt off 1pt,color=wqwqwq] (45.24725627003013,12.266150228521623)-- (45.23502077410504,10.866251421990667);
\draw [line width=0.4pt,dash pattern=on 1pt off 1pt,color=wqwqwq] (45.9967931144536,9.371436149925437)-- (46.00043486637473,13.620022275016726);
\draw [line width=0.4pt,dash pattern=on 1pt off 1pt,color=wqwqwq] (47.164998691032324,9.233797044949897)-- (47.231982489608505,13.480409364915651);
\draw [line width=0.4pt,dash pattern=on 1pt off 1pt,color=wqwqwq] (44.61255113911584,9.534528871253691)-- (42.71752206031173,8.744486338211138);
\draw [line width=0.4pt,dash pattern=on 1pt off 1pt,color=wqwqwq] (46.629706887503126,12.125799912026897)-- (46.614079363928,13.550457207085401);
\draw [line width=0.4pt,dash pattern=on 1pt off 1pt,color=wqwqwq] (42.9909004526187,13.417809288014508)-- (43.003654179879916,9.227857905781066);
\draw [line width=0.4pt,dash pattern=on 1pt off 1pt,color=wqwqwq] (43.00365417987992,9.227857905781072)-- (46.86798651724574,8.72432554077885);
\draw [line width=0.4pt,dash pattern=on 1pt off 1pt,color=wqwqwq] (43.91693314489181,9.61648725259465)-- (44.00655503164835,13.846056056306544);
\draw [line width=2.pt,color=qqqqff] (45.996444258330776,8.339859280010607)-- (47.84206589952466,9.154024335031758);
\begin{scriptsize}
\draw [fill=qqffqq] (42.00099901480501,11.65020176655288) circle (3.5pt);
\draw [fill=qqffqq] (47.852470594694445,12.001661464596303) circle (3.5pt);
\draw [fill=ffqqqq] (47.84206589952466,10.583917241350086) circle (3.5pt);
\draw [fill=qqffqq] (43.41491911559936,10.075034968866147) circle (3.5pt);
\draw [fill=qqffqq] (45.188280748486896,9.46749440945095) circle (3.5pt);
\draw [fill=qqffqq] (46.56756201851053,9.303294258257653) circle (3.5pt);
\draw [fill=qqffqq] (47.84206589952466,9.154024335031758) circle (3.5pt);
\draw [fill=qqffqq] (45.996444258330776,8.339859280010607) circle (3.5pt);
\draw [fill=ffqqqq] (45.24725627003013,12.266150228521623) circle (3.5pt);
\draw [fill=qqffqq] (45.26552966374888,13.704257400946888) circle (3.5pt);
\draw [fill=ffqqqq] (43.38255561863366,12.91052541396301) circle (3.5pt);
\draw [fill=qqffqq] (42.,13.) circle (3.5pt);
\draw [fill=qqffqq] (45.99749611216069,11.450219140469242) circle (3.5pt);
\draw [fill=ffqqqq] (46.629706887503126,12.125799912026897) circle (3.5pt);
\draw [fill=ffqqqq] (45.23502077410504,10.866251421990667) circle (3.5pt);
\draw [fill=ffqqqq] (44.61358021931038,9.94367484791151) circle (3.5pt);
\draw [fill=ffqqqq] (45.99694230609766,9.812599824913848) circle (3.5pt);
\draw [fill=ffqqqq] (46.59872296377298,10.718567081353992) circle (3.5pt);
\draw [fill=ffqqqq] (43.39773833016538,11.5803096611827) circle (3.5pt);
\draw [fill=ffqqqq] (44.609474050822676,11.519675039004012) circle (3.5pt);
\end{scriptsize}
\end{tikzpicture}
\caption{An optimal conformation for the H-P protein sequence $s$.}
\label{figure1}
\end{figure}
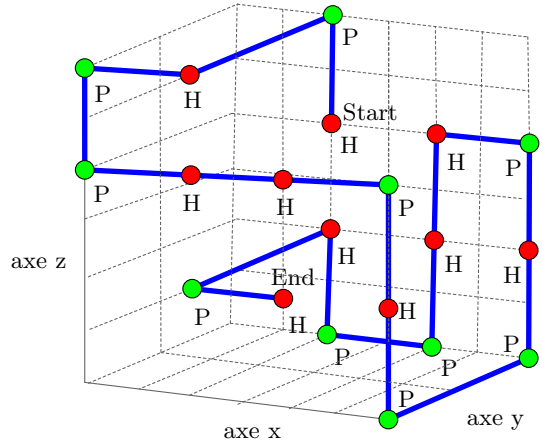

In the HP lattice model, the twenty amino acids  have been classified into two classes: the Hydrophobic (H) class and the Polar (P) class. For any given protein sequence that contains $n$ amino acids, the primary sequence is abstracted into another sequence $s=s_{1}s_{2}\ldots s_{n}$, with $s_{i} \in \mbox{\{H,P\}}$ for $i=1,2,\ldots,n$, so that a feasible conformation for $s$ is presented by a self-evident walk in the lattice, and the quality of folding (i.e., free energy $E(s)$) is measured by the number of H-H topological contacts \cite{lin_efficient_2009}. Furthermore, any pair of amino acids forms a contact if the two amino acids are located in two adjacent nodes of the lattice and are not consecutive in the primary sequence \cite{lau_lattice_1989}. Each H-H topological contact decreases the free energy by 1 \mbox{(i.e., H-H $ \equiv $ -1, and P-P $ \equiv P-H \equiv 0 $) \cite{lin_efficient_2009}.}
\begin{equation}\label{1}
   E(s)=-\sum_{i=1}^{n-2}\sum_{j=i+2}^{n}h_{ij}x_{ij}
 \end{equation}
 Such as:
\begin{equation}\label{2}
h_{ij}=\left\{
  \begin{array}{ll}
    1, & \hbox{if $s_{i}=s_{j}=\textrm{H}$}, \\
    0, & \hbox{otherwise}.
  \end{array}
\right.
\end{equation}
And
\begin{equation}\label{3}
x_{ij}=\left\{
  \begin{array}{ll}
    1, & \hbox{If the amino acids $i$ and $j$ form a topological contact,} \\
    0, & \hbox{otherwise}.
  \end{array}
\right.
\end{equation}
According to the simplified HP model and the definition of the  energy function in this model, the PFP can be expressed as an optimization problem as follows: let  $s=s_{1}s_{2}\ldots s_{n}$ be a protein sequence of $n$ amino acids in the HP model. The challenge is to identify a feasible conformation $ c^{*}(s)\in C $ such that $E(c^{*}(s))=min\{ E(c(s))\mid c(s)\in C\}$, where $C$ is the space of conformations \cite{shmygelska_ant_2005}. As an example, in Figure \ref{figure1}, we present an optimal conformation in the 3D-HP cubic lattice model for the HP sequence $ s=HPHPPHPHPHPHPHPHPHPHPHPHPHPHPHPHPH $, which has 11 H-H topological contacts.  Hence, the free energy of this conformation is \mbox{$E(c(s))=-11$}.
\section{Mathematical Formulation for the PFP Problem }
This section offers a   mathematical formulation with quadratic objective function for the protein structure prediction problem in its 3D cubic lattice model. A similar formulation has been proposed for the 2D triangular lattice model in \cite{sadek_new_2021-1}. In the 3D cubic lattice, each node has six neighbors. Before we present the suggested mathematical formulation for the PFP problem, we consider the following encoding neighbors for any given position $(i,j,k)$ in the 3D cubic lattice.
  \begin{figure}[H]
\begin{center}
    \includegraphics[width=7cm,height=7cm]{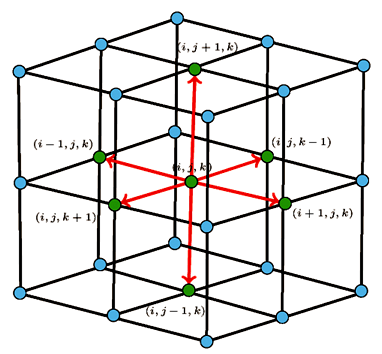}
\end{center}
  \caption{ The six adjacent positions to the  position $(i,j,k)$ in the 3D cubic lattice.}
\label{figure2}
\end{figure}

Suppose that $n$ is the sequence length of the protein, and $x_{ijk}^{l}$ is a binary decision variable such that: \vspace{0.2em}
$$x_{ijk}^{l} = \left\{
\begin{array}{ll}
1, & \textrm{if the position }\ (i,j,k)\ \textrm{contain the}\ l^{th}\ \textrm{amino acid in the protien  sequence},\vspace{0.2cm}\\
0, & \textrm{else}.
\end{array}
\right.
$$
\subsection{Constraints}
First, we assign the first amino acid of the sequence at position $(n,n,n)$ to be the starting point, i.e., \vspace{0.3em} $$x_{nnn}^{1}=1.$$ \vspace{0.3em}
In terms of constraints, we can identify three main constraints that ensure the admissibility of the produced solution:
\begin{itemize}
\item[1.] A feasible conformation  must occupies  $n$ nodes in the lattice. This constraint can be formulated  as follows:\vspace{0.3em}
$$ \sum_{l=1}^{n}\sum_{i=1}^{2n}\sum_{j=1}^{2n}\sum_{k=1}^{2n}x_{ijk}^{l} = n.$$\vspace{0.3em}
\item[2.] Any node in the lattice may have at maximum one amino acid, so :\vspace{0.3em}
$$ \sum_{l=1}^{n}x_{ijk}^{l} \leq 1,\ \forall i \in \{1,\dots, 2n\},\ \forall j \in \{1,\dots, 2n\}, \forall k \in \{1,\dots, 2n\}. $$\vspace{0.3em}
\item[3.] A given node of the lattice can store the amino acid at position $l+1$, if at least one of its neighbor's node has the $l^{th}$ amino acid:\vspace{0.3em}
$$x_{ijk}^{l+1} \leq  x_{i-1jk}^{l}+x_{i+1jk}^{l}+x_{ij-1k}^{l}+x_{ij+1k}^{l}+x_{ijk-1}^{l}+x_{ijk+1}^{l},\forall i,j,k \in \{1,\dots, 2n\},\ \forall l \in \{1,\dots,n-1\}.$$\vspace{0.3em}
\end{itemize}
\subsection{The objective function}
Let $\lambda_{l}$ be a binary representation of the protein sequence (i.e., a $\{0,1\}$-representation) which is based on the H-P model classification of amino acids, where :\vspace{0.3em}
$$\lambda_{l} = \left\{
\begin{array}{ll}
1 & \mbox{if the  $l^{th}$ amino acid in the protein  sequence  is hydrophobic, i.e., H,}\vspace{0.2cm}\\
0 & \mbox{if the  $l^{th}$ amino acid in the  protein sequence  is hydrophilic , i.e., P.}
\end{array}
\right.\vspace{0.4em}
$$
Therefore, the objective function can be expressed by the following quadratic formula:\vspace{0.3em}
$$\max(\mathcal{Z})=\frac{1}{2}\mathcal{Z^{*}}-\sum_{l=1}^{n-1}\lambda_{l}\lambda_{l+1},$$ \vspace{0.3em}
where: \vspace{0.4em}
\begin{small}
$$\hspace{-2mm}\mathcal{Z^{*}}=\max_{x}\left\{\sum_{i=1}^{2n}\sum_{j=1}^{2n}\sum_{k=1}^{2n}\left(\sum_{l=1}^{n}\lambda_{l}x_{ijk}^{l}\right)\left(\sum_{l=1}^{n}\lambda_{l}\left(x_{i-1jk}^{l}+x_{i+1jk}^{l}+x_{ij-1k}^{l}+x_{ij+1k}^{l}+x_{ijk-1}^{l}+x_{ijk+1}^{l}\right)\right)\right\}.$$
\end{small}\vspace{0.5em}
This mathematical modeling allows to find an optimal solution which is included in the lattice bounded by the points $\{(1,1,1),\ldots,(2n,2n,2n)\}$, with starting point $(n,n,n)$.
\section{Cuckoo Search Algorithm}
The Cuckoo Search Algorithm (CSA) is one of the most recently developed metaheuristics, which was first proposed in 2009 by Yang and Deb \cite{yang_cuckoo_2009}. \\
\noindent\begin{minipage}{11.65cm}
These algorithm models the natural strategy of obligate brood parasitism of a particular species called cuckoo. Cuckoo species make the transition from the current generation to the next by laying their eggs in the nests of other species. Nevertheless, a number of host birds may discover a direct conflict with the introduced cuckoo eggs. In this case, two scenarios are possible: if a host bird identifies any cuckoo eggs, it will either throw the eggs out of its nest or abandon its nest and construct a new one.
\end{minipage}
\begin{minipage}{5cm}
  \begin{tabular}{c}\vspace{-0.4cm}
    \includegraphics[width=4cm,height=4cm]{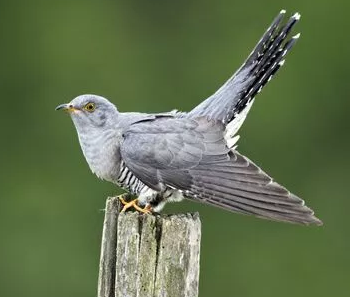}\vspace{0.2cm}\\
    \footnotesize{Cuckoo Bird}
  \end{tabular}
\end{minipage}

\vspace{0.1cm}\noindent  To use this strategy to solve optimization problems, Yang and Deb proposed the following \mbox{rules \cite{yang_cuckoo_2009}:}
\begin{enumerate}
  \item Each cuckoo lays only one egg at a time in a randomly selected nest.
  \item The nest with high-quality eggs (i.e., the solutions with the best fitness) will pass to the next generation.
  \item The number of host nests is fixed, and the host bird can discover a cuckoo egg with a probability $p_{c} \in [0,1]$. In this case,  the host bird can either throw the egg away or abandon the nest and rebuild a new nest in another site.
\end{enumerate}
The CSA pseudo-code is presented in the following algorithm:
\begin{algorithm}[htp]
\setstretch{1}
\caption{Cuckoo Search Algorithm: CSA}
\begin{flushleft}
\algorithmicrequire  Problem instance $I$.\vspace{0.1cm} \\
\algorithmicensure  The best-found solution for $I$.\\ \vspace{-0.29cm}
\rule{\linewidth}{0.02pt}
\end{flushleft}
\begin{algorithmic}[H]
\State\hspace{-7mm} \textbf{Begin}
 \State   Generate initial population $P$ of $m$ solutions ($m$ host nests) $x=(x_{1},\ldots,x_{m})$;
 \State  $iter=0$;
 \While {$iter\leq$ Max (maximum number of generations)}
  \State  $i = 1 $;
  \While {$i \leq m$}
\State Get a randomly cuckoo (say $y_{i}$) via L\'{e}vy Flights and evaluate its fitness $f(y_{i})$.
\State Select one  nest among $m$ randomly (say $x_{i}$).
\If {$f(x_{i})> f(y_{i})$}
\State Replace $x_{i}$ by $y_{i}$ ;
\EndIf \State \textbf{end if}
\State $i=i+1;$
\EndWhile
\State \textbf{end while}
\State  A fraction $r_{a}$ of worse nest are abandoned  by host birds, and a new ones are built;
\State Rank the solutions and find the best one.
\State Update the current optimal solution.
\State $iter=iter+1;$
\EndWhile
\State \textbf{end while}
\State\hspace{-7mm}  \textbf{End}\vspace{2mm}
\end{algorithmic}
\label{Algorithm1}
\end{algorithm}\\
We notice that a cuckoo corresponds to a solution in the search space and that a cuckoo egg corresponds to a new solution. Moreover, for simplicity, each nest has only one egg. The quality of nests and cuckoo eggs is measured by the objective function, and if a cuckoo egg is discovered by the host bird, it rebuilds a new nest at another site with a new random solution.
The first step of CSA is to generate the initial nests (the initial population). Then, for each generation $t$, a new solution $ X_{i}^{t+1}$ is generated for each cuckoo $X_{i}^{t}$ by a random walk via L\'{e}vy Flights:
\vspace{0.4em}
\begin{equation}\label{4}
  X_{i}^{t+1}= X_{i}^{t}+ \alpha\oplus \textrm{l\'{e}vy}(\gamma),
\end{equation}
\vspace{0.4em}
where $\alpha > 0$ is the step size witch is depends on the the interesting  optimization problem interests (in most cases  $\alpha =1$), the operator $\oplus$ means entry-wise multiplications  and  $\textrm{l\'{e}vy}(\gamma)$ is the step length which distributed according to the l\'{e}vy distribution: \vspace{0.4em}
\begin{equation}\label{4}
\textrm{l\'{e}vy}(\gamma) \rightsquigarrow u= z^{-\gamma},\ (1< \gamma \leq 3).
\end{equation}
\vspace{0.4em}
 CSA combined with L\'{e}vy Flights was originally designed to deal with continuous  optimization  problems. The simulation results presented in \cite{yang_cuckoo_2009}, have demonstrated that CSA is more efficient than GA and PSO algorithms for multimodal objective functions. With its high exploration capacity of the search space and its small number of parameters when compared to other algorithms, CSA has been widely used to address several continuous optimization problems \cite{yang_cuckoo_2009, ouaarab_discrete_2014, lim_hybrid_2016}.
 In this paper, we propose a discrete hybrid cuckoo search algorithm and we use it to solve the protein folding problem which is one of the most difficult combinatorial optimization problems.
\section{Hybrid Cuckoo Search Algorithm (HCSA) for PFP}
In order to solve the PFP problem, in this paper we suggest a Hybrid Cuckoo Search Algorithm, namely HCSA, which combines the cuckoo search algorithm with the Hill-Climbing algorithm and the crossover operator of GA. The idea behind this hybridization is to provide a good balance between exploration and exploitation for the search process. Both the crossover operator and HC algorithm are effective techniques for the exploitation of solutions. While L\'{e}vy Flights is a very effective technique in the diversification phase through its capacity to explore new regions in the search space. The pseudo-code of the proposed algorithm (HCSA) is illustrated in Algorithm \ref{Algorithm2}.
\subsection{Representation of Solutions and Initial Population}
HCSA algorithm starts with a set of $m$ nests (i.e., the initial population), where each nest represents a randomly generated solution in the decision space. Given $n$ the number of amino acids in a given protein sequence $s$, a feasible solution for $s$ is encoded by a sequence $c$ of $(n-1)$ directions of movement in the 3D cubic lattice $c\in {\{R, L, U, D, F, B\}}^{n-1}$, where the symbols $R, L, U, D, F, B$ refer to directions: Right, Left, Up, Down, Forward and Backward respectively as illustrated in Figure \ref{figure3}.
  \begin{figure}[H]
\begin{center}
    \includegraphics[width=5cm,height=5cm]{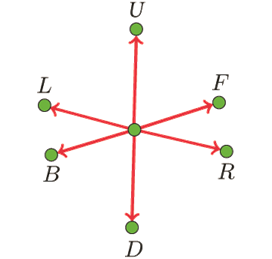}
\end{center}
\caption{Movement directions in the 3D cubic lattice.}
\label{figure3}
\end{figure}
\begin{algorithm}[H]
\setstretch{1}
\caption{The Suggested hybrid algorithm: HCSA}
\begin{flushleft}
\algorithmicrequire  Problem instance(i.e., protein sequence)  $I$.\vspace{0.1cm} \\
\algorithmicensure  The best-found conformation for $I$.\\ \vspace{-0.29cm}
\rule{\linewidth}{0.02pt}
\end{flushleft}
\begin{algorithmic}
\State\hspace{-7mm} \textbf{Begin}
 \State   Generate initial population $P$ of $m$ solutions ($m$ host nests) $x=(x_{1},\ldots,x_{m})$;
 \State  $iter=0$;
 \While {$iter\leq Max$ (maximum number of generations)}
  \State  $i = 1 $;
  \While {$i \leq Max$}
 \State Select a  solution $x_{j}$ from $P$ using the selection operator;
\State Produce a  new cuckoo eggs $y_{i}$ by applying the crossover operator to the selected
\State  parents $x_{i}$ and $x_{j}$;
\State Improve the quality of $y_{i}$ using the Hill-Climbing Algorithm;
\If {$f(y_{i})> f(x_{i})$}
\State Replace $x_{i}$ by $y_{i}$ ;
\EndIf \State \textbf{end if}
\State $i=i+1;$
\EndWhile
\State \textbf{end while}
\State The worse nest is discarded with probability $p_{a} \in(0,1)$ by host birds, and a new nest is
\State generated via L\'{e}vy Flights.
\State  Rotation of represented solutions with probability $p_{c} \in(0,1)$.
\State Rank the solutions and find the best one.
\State Update the current optimal solution;
\State $iter=iter+1;$
\EndWhile
\State \textbf{end while}
\State\hspace{-7mm}  \textbf{End}\vspace{2mm}
\end{algorithmic}
\label{Algorithm2}
\end{algorithm}

\subsection{Generation of a New Cuckoo Egg by  the Crossover Operator}
The crossover operator is one of the genetic operators that apply to some selected solutions (called parents) to create new solutions (called offspring) for the next generation. Here, we apply a random 1-point crossover $h\in\{2,3,\ldots,n-1\}$. For each solution from population $P$ that we consider as the first parent, we randomly select another one from the current population as the second parent. We then apply the crossover operation. Figure \ref{figure3} shows two offspring that are produced by combining (i.e., swapping the site directions) the two selected parents at the point $h$. In the suggested algorithm only the best offspring is considered for the next step.
\begin{figure}[htp]
\definecolor{qqqqff}{rgb}{0.,0.,1.}
\definecolor{qqffqq}{rgb}{0.,1.,0.}
\definecolor{ffqqqq}{rgb}{1.,0.,0.}
\definecolor{cqcqcq}{rgb}{0.7529411764705882,0.7529411764705882,0.7529411764705882}
\begin{tikzpicture}[transform shape, scale = 1.05]
\draw [color=cqcqcq,, xstep=0.5cm,ystep=0.5cm] (18.914810658298833,8.945260273243697) grid (26.563145510156314,16.55530558892144);
\clip(18.914810658298833,8.945260273243697) rectangle (26.563145510156314,16.55530558892144);
\draw (20.28,13.18) node[anchor=north west] {\footnotesize{Parent 1}};
\draw (24,13.18) node[anchor=north west] {\footnotesize{Parent 2}};
\draw (20.33,9.68) node[anchor=north west] {\footnotesize{Offspring 1}};
\draw (23.53,9.68) node[anchor=north west] {\footnotesize{Offspring 2}};
\draw (21.5,16.15) node[anchor=north west] {\footnotesize{Crossover point}};
\draw [line width=1.2pt,color=qqqqff] (20.5,15.)-- (20.5,15.5);
\draw [line width=1.2pt,color=qqqqff] (20.5,15.5)-- (21.,15.5);
\draw [line width=1.2pt,color=qqqqff] (21.,15.5)-- (21.,15.);
\draw [line width=1.2pt,color=qqqqff] (21.,15.)-- (21.5,15.);
\draw [line width=1.2pt,color=qqqqff] (21.5,15.)-- (22.,15.);
\draw [line width=1.2pt,color=qqqqff] (22.,15.)-- (22.,14.5);
\draw [line width=1.2pt,color=qqqqff] (22.,14.5)-- (21.5,14.5);
\draw [line width=1.2pt,color=qqqqff] (21.5,14.5)-- (21.5,14.);
\draw [line width=1.2pt,color=qqqqff] (21.5,14.)-- (21.,14.);
\draw [line width=1.2pt,color=qqqqff] (21.,14.)-- (21.,13.5);
\draw [line width=1.2pt,color=qqqqff] (21.,13.5)-- (20.5,13.5);
\draw [line width=1.2pt,color=qqqqff] (20.5,13.5)-- (20.,13.5);
\draw [line width=1.2pt,color=qqqqff] (20.,13.5)-- (20.,14.);
\draw [line width=1.2pt,color=qqqqff] (20.,14.)-- (20.,14.5);
\draw [line width=1.2pt,color=qqqqff] (20.,14.5)-- (20.5,14.5);
\draw [->] (22.712639397772925,15.63191642659372) -- (22.135877034132335,14.572942906466759);
\draw [line width=1.2pt,color=qqqqff] (20.502497668942326,11.490607142782093)-- (20.502497668942326,11.990607142782093);
\draw [line width=1.2pt,color=qqqqff] (20.502497668942326,11.990607142782093)-- (21.002497668942326,11.990607142782093);
\draw [line width=1.2pt,color=qqqqff] (21.002497668942326,11.990607142782093)-- (21.002497668942326,11.490607142782093);
\draw [line width=1.2pt,color=qqqqff] (21.002497668942326,11.490607142782093)-- (21.502497668942326,11.490607142782093);
\draw [line width=1.2pt,color=qqqqff] (21.502497668942326,11.490607142782093)-- (22.002497668942326,11.490607142782093);
\draw [line width=1.2pt,color=qqqqff] (22.002497668942326,11.490607142782093)-- (22.002497668942326,10.990607142782093);
\draw [line width=1.2pt,color=qqqqff] (22.002497668942326,10.990607142782093)-- (21.502497668942326,10.990607142782093);
\draw [line width=1.2pt,color=qqqqff] (21.502497668942326,10.990607142782093)-- (21.002497668942326,10.990607142782093);
\draw [line width=1.2pt,color=qqqqff] (21.002497668942326,10.990607142782093)-- (21.002497668942326,10.490607142782093);
\draw [line width=1.2pt,color=qqqqff] (21.002497668942326,10.490607142782093)-- (21.502497668942326,10.490607142782093);
\draw [line width=1.2pt,color=qqqqff] (21.502497668942326,10.490607142782093)-- (21.502497668942326,9.990607142782093);
\draw [line width=1.2pt,color=qqqqff] (21.502497668942326,9.990607142782093)-- (21.002497668942326,9.990607142782093);
\draw [line width=1.2pt,color=qqqqff] (21.002497668942326,9.990607142782093)-- (20.502497668942326,9.990607142782093);
\draw [line width=1.2pt,color=qqqqff] (20.502497668942326,9.990607142782093)-- (20.502497668942326,10.490607142782093);
\draw [line width=1.2pt,color=qqqqff] (20.502497668942326,10.490607142782093)-- (20.502497668942326,10.990607142782093);
\draw [line width=1.2pt,color=qqqqff] (24.00044234327216,14.987543914502325)-- (24.00044234327216,15.487543914502325);
\draw [line width=1.2pt,color=qqqqff] (24.00044234327216,15.487543914502325)-- (24.50044234327216,15.487543914502325);
\draw [line width=1.2pt,color=qqqqff] (24.50044234327216,15.487543914502325)-- (25.00044234327216,15.487543914502325);
\draw [line width=1.2pt,color=qqqqff] (25.00044234327216,15.487543914502325)-- (25.50044234327216,15.487543914502325);
\draw [line width=1.2pt,color=qqqqff] (25.50044234327216,15.487543914502325)-- (25.50044234327216,14.987543914502325);
\draw [line width=1.2pt,color=qqqqff] (25.50044234327216,14.987543914502325)-- (25.50044234327216,14.487543914502325);
\draw [line width=1.2pt,color=qqqqff] (25.50044234327216,14.487543914502325)-- (25.00044234327216,14.487543914502325);
\draw [line width=1.2pt,color=qqqqff] (25.00044234327216,14.487543914502325)-- (24.50044234327216,14.487543914502325);
\draw [line width=1.2pt,color=qqqqff] (24.50044234327216,14.487543914502325)-- (24.50044234327216,13.987543914502325);
\draw [line width=1.2pt,color=qqqqff] (24.50044234327216,13.987543914502325)-- (25.00044234327216,13.987543914502325);
\draw [line width=1.2pt,color=qqqqff] (25.00044234327216,13.987543914502325)-- (25.00044234327216,13.487543914502325);
\draw [line width=1.2pt,color=qqqqff] (25.00044234327216,13.487543914502325)-- (24.50044234327216,13.487543914502325);
\draw [line width=1.2pt,color=qqqqff] (24.50044234327216,13.487543914502325)-- (24.00044234327216,13.487543914502325);
\draw [line width=1.2pt,color=qqqqff] (24.00044234327216,13.487543914502325)-- (24.00044234327216,13.987543914502325);
\draw [line width=1.2pt,color=qqqqff] (24.00044234327216,13.987543914502325)-- (24.00044234327216,14.487543914502325);
\draw [line width=1.2pt,color=qqqqff] (23.50780142753228,10.990607142782093)-- (23.50780142753228,10.490607142782093);
\draw [line width=1.2pt,color=qqqqff] (23.50780142753228,10.490607142782093)-- (23.50780142753228,9.990607142782093);
\draw [line width=1.2pt,color=qqqqff] (23.50780142753228,9.990607142782093)-- (24.00780142753228,9.990607142782093);
\draw [line width=1.2pt,color=qqqqff] (24.00780142753228,9.990607142782093)-- (24.50780142753228,9.990607142782093);
\draw [line width=1.2pt,color=qqqqff] (24.50780142753228,9.990607142782093)-- (24.50780142753228,10.490607142782093);
\draw [line width=1.2pt,color=qqqqff] (24.50780142753228,10.490607142782093)-- (25.00780142753228,10.490607142782093);
\draw [line width=1.2pt,color=qqqqff] (25.00780142753228,10.490607142782093)-- (25.00780142753228,10.990607142782093);
\draw [line width=1.2pt,color=qqqqff] (25.00780142753228,10.990607142782093)-- (25.50780142753228,10.990607142782093);
\draw [line width=1.2pt,color=qqqqff] (25.50780142753228,10.990607142782093)-- (25.50780142753228,11.490607142782093);
\draw [line width=1.2pt,color=qqqqff] (25.50780142753228,11.490607142782093)-- (25.50780142753228,11.990607142782093);
\draw [line width=1.2pt,color=qqqqff] (25.50780142753228,11.990607142782093)-- (25.00780142753228,11.990607142782093);
\draw [line width=1.2pt,color=qqqqff] (25.00780142753228,11.990607142782093)-- (24.50780142753228,11.990607142782093);
\draw [line width=1.2pt,color=qqqqff] (24.50780142753228,11.990607142782093)-- (24.00780142753228,11.990607142782093);
\draw [line width=1.2pt,color=qqqqff] (24.00780142753228,11.990607142782093)-- (24.00780142753228,11.490607142782093);
\draw [line width=1.2pt,color=qqqqff] (23.50780142753228,10.990607142782093)-- (24.00780142753228,10.990607142782093);
\begin{scriptsize}
\draw [fill=ffqqqq] (20.5,15.) circle (3.0pt);
\draw [fill=qqffqq] (20.5,15.5) circle (3.0pt);
\draw [fill=qqffqq] (21.,15.5) circle (3.0pt);
\draw [fill=ffqqqq] (21.,15.) circle (3.0pt);
\draw [fill=ffqqqq] (21.5,15.) circle (3.0pt);
\draw [fill=qqffqq] (22.,15.) circle (3.0pt);
\draw [fill=qqffqq] (22.,14.5) circle (3.0pt);
\draw [fill=ffqqqq] (21.5,14.5) circle (3.0pt);
\draw [fill=ffqqqq] (21.5,14.) circle (3.0pt);
\draw [fill=ffqqqq] (21.,14.) circle (3.0pt);
\draw [fill=qqffqq] (21.,13.5) circle (3.0pt);
\draw [fill=qqffqq] (20.5,13.5) circle (3.0pt);
\draw [fill=ffqqqq] (20.,13.5) circle (3.0pt);
\draw [fill=qqffqq] (20.,14.) circle (3.0pt);
\draw [fill=ffqqqq] (20.,14.5) circle (3.0pt);
\draw [fill=ffqqqq] (20.5,14.5) circle (3.0pt);
\draw [fill=ffqqqq] (20.502497668942326,11.490607142782093) circle (3.0pt);
\draw [fill=qqffqq] (20.502497668942326,11.990607142782093) circle (3.0pt);
\draw [fill=qqffqq] (21.002497668942326,11.990607142782093) circle (3.0pt);
\draw [fill=ffqqqq] (21.002497668942326,11.490607142782093) circle (3.0pt);
\draw [fill=ffqqqq] (21.502497668942326,11.490607142782093) circle (3.0pt);
\draw [fill=qqffqq] (22.002497668942326,11.490607142782093) circle (3.0pt);
\draw [fill=qqffqq] (22.002497668942326,10.990607142782093) circle (3.0pt);
\draw [fill=ffqqqq] (21.502497668942326,10.990607142782093) circle (3.0pt);
\draw [fill=ffqqqq] (21.002497668942326,10.990607142782093) circle (3.0pt);
\draw [fill=ffqqqq] (21.002497668942326,10.490607142782093) circle (3.0pt);
\draw [fill=qqffqq] (21.502497668942326,10.490607142782093) circle (3.0pt);
\draw [fill=qqffqq] (21.502497668942326,9.990607142782093) circle (3.0pt);
\draw [fill=ffqqqq] (21.002497668942326,9.990607142782093) circle (3.0pt);
\draw [fill=qqffqq] (20.502497668942326,9.990607142782093) circle (3.0pt);
\draw [fill=ffqqqq] (20.502497668942326,10.490607142782093) circle (3.0pt);
\draw [fill=ffqqqq] (20.502497668942326,10.990607142782093) circle (3.0pt);
\draw [fill=ffqqqq] (24.00044234327216,14.987543914502325) circle (3.0pt);
\draw [fill=qqffqq] (24.00044234327216,15.487543914502325) circle (3.0pt);
\draw [fill=qqffqq] (24.50044234327216,15.487543914502325) circle (3.0pt);
\draw [fill=ffqqqq] (25.00044234327216,15.487543914502325) circle (3.0pt);
\draw [fill=ffqqqq] (25.50044234327216,15.487543914502325) circle (3.0pt);
\draw [fill=qqffqq] (25.50044234327216,14.987543914502325) circle (3.0pt);
\draw [fill=qqffqq] (25.50044234327216,14.487543914502325) circle (3.0pt);
\draw [fill=ffqqqq] (25.00044234327216,14.487543914502325) circle (3.0pt);
\draw [fill=qqffqq] (24.50044234327216,14.487543914502325) circle (3.0pt);
\draw [fill=ffqqqq] (24.50044234327216,13.987543914502325) circle (3.0pt);
\draw [fill=qqffqq] (25.00044234327216,13.987543914502325) circle (3.0pt);
\draw [fill=qqffqq] (25.00044234327216,13.487543914502325) circle (3.0pt);
\draw [fill=ffqqqq] (24.50044234327216,13.487543914502325) circle (3.0pt);
\draw [fill=qqffqq] (24.00044234327216,13.487543914502325) circle (3.0pt);
\draw [fill=qqffqq] (24.00044234327216,13.987543914502325) circle (3.0pt);
\draw [fill=ffqqqq] (24.00044234327216,14.487543914502325) circle (3.0pt);
\draw [fill=ffqqqq] (23.50780142753228,10.990607142782093) circle (3.0pt);
\draw [fill=qqffqq] (23.50780142753228,10.490607142782093) circle (3.0pt);
\draw [fill=ffqqqq] (23.50780142753228,9.990607142782093) circle (3.0pt);
\draw [fill=qqffqq] (24.00780142753228,9.990607142782093) circle (3.0pt);
\draw [fill=qqffqq] (24.50780142753228,9.990607142782093) circle (3.0pt);
\draw [fill=ffqqqq] (24.50780142753228,10.490607142782093) circle (3.0pt);
\draw [fill=ffqqqq] (25.00780142753228,10.490607142782093) circle (3.0pt);
\draw [fill=ffqqqq] (25.00780142753228,10.990607142782093) circle (3.0pt);
\draw [fill=qqffqq] (25.50780142753228,10.990607142782093) circle (3.0pt);
\draw [fill=qqffqq] (25.50780142753228,11.490607142782093) circle (3.0pt);
\draw [fill=ffqqqq] (25.50780142753228,11.990607142782093) circle (3.0pt);
\draw [fill=ffqqqq] (25.00780142753228,11.990607142782093) circle (3.0pt);
\draw [fill=qqffqq] (24.50780142753228,11.990607142782093) circle (3.0pt);
\draw [fill=qqffqq] (24.00780142753228,11.990607142782093) circle (3.0pt);
\draw [fill=ffqqqq] (24.00780142753228,11.490607142782093) circle (3.0pt);
\draw [fill=ffqqqq] (24.00780142753228,10.990607142782093) circle (3.0pt);
\end{scriptsize}
\end{tikzpicture}
\caption{An illustrative example of 1-point crossover operator.}
\label{figure3}
\end{figure}
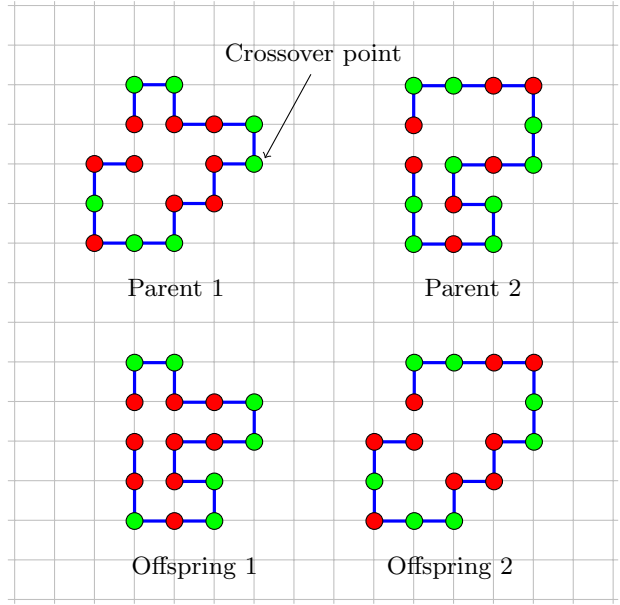
\subsection{Improvement of the Quality of Eggs by the Hill Claiming Algorithm}
To enhance the quality of the eggs that are produced using the crossover operator, we input each of them as an initial solution into the  Hill Claiming Algorithm (HCA). The proposed HCA starting from a solution $s$ and improving its quality by exploiting its neighbors $N(s)$.  This algorithm uses, as a search strategy, the diagonal pull move to explore the neighborhood of a given solution \cite{bockenhauer_local_2008}. For each iteration, we select an amino acid and we replace its position in the lattice with one of its free neighborhood positions, if it exists, as we show in Figure 4. Moreover, the worst-case move can be accepted according to a random walk acceptance criterion (i.e., the $\beta$ operator). This exploration technique minimizes the risk of premature convergence to a local optimum. Then, we evaluate the new resulting solution $s'$, and we compare its quality with the current solution; if $f(s')< f(s)$, or a random value $ rand < \beta$, we replace $s$ by $s'$. HCA is summarized in the algorithm \ref{Algorithm3}.
  \begin{center}
\begin{figure}[H]
\definecolor{qqffqq}{rgb}{0.,1.,0.}
\definecolor{qqqqff}{rgb}{0.,0.,1.}
\definecolor{ffqqqq}{rgb}{1.,0.,0.}
\definecolor{cqcqcq}{rgb}{0.7529411764705882,0.7529411764705882,0.7529411764705882}
\begin{tikzpicture}[transform shape, scale = 1.05]
\draw [color=cqcqcq,, xstep=0.5cm,ystep=0.5cm] (18.92438304234371,12.448752833668921) grid (26.055809155777784,16.057541618587802);
\clip(18.92438304234371,12.448752833668921) rectangle (26.055809155777784,16.057541618587802);
\draw [line width=1.2pt,color=qqqqff] (21.,14.5)-- (20.5,14.5);
\draw [line width=1.2pt,color=qqqqff] (20.5,14.5)-- (20.5,15.);
\draw [line width=1.2pt,color=qqqqff] (20.5,15.)-- (21.,15.);
\draw [line width=1.2pt,color=qqqqff] (21.,15.)-- (21.5,15.);
\draw [line width=1.2pt,color=qqqqff] (21.5,15.)-- (21.5,14.5);
\draw [line width=1.2pt,color=qqqqff] (21.5,14.5)-- (21.5,14.);
\draw [line width=1.2pt,color=qqqqff] (21.5,14.)-- (21.5,13.5);
\draw [line width=1.2pt,color=qqqqff] (21.5,13.5)-- (21.,13.5);
\draw [line width=1.2pt,color=qqqqff] (21.,13.5)-- (20.5,13.5);
\draw [line width=1.2pt,color=qqqqff] (20.5,13.5)-- (20.,13.5);
\draw [line width=1.2pt,color=qqqqff] (20.,13.5)-- (20.,14.);
\draw [line width=1.2pt,color=qqqqff] (20.,14.)-- (20.,14.5);
\draw [line width=1.2pt,color=qqqqff] (20.,14.5)-- (20.,15.);
\draw [line width=1.2pt,color=qqqqff] (24.5,14.5)-- (24.,14.5);
\draw [line width=1.2pt,color=qqqqff] (24.,14.5)-- (24.,15.);
\draw [line width=1.2pt,color=qqqqff] (24.,15.)-- (24.5,15.);
\draw [line width=1.2pt,color=qqqqff] (24.5,15.)-- (25.,15.);
\draw [line width=1.2pt,color=qqqqff] (25.,15.)-- (25.00044234327216,14.487543914502325);
\draw [line width=1.2pt,color=qqqqff] (25.00044234327216,14.487543914502325)-- (25.,14.);
\draw [line width=1.2pt,color=qqqqff] (25.,14.)-- (24.50044234327216,13.987543914502325);
\draw [line width=1.2pt,color=qqqqff] (24.50044234327216,13.987543914502325)-- (24.5,13.5);
\draw [line width=1.2pt,color=qqqqff] (24.5,13.5)-- (24.,13.5);
\draw [line width=1.2pt,color=qqqqff] (24.,13.5)-- (24.,14.);
\draw [line width=1.2pt,color=qqqqff] (24.,14.)-- (23.5,14.);
\draw [line width=1.2pt,color=qqqqff] (23.5,14.)-- (23.5,14.5);
\draw [line width=1.2pt,color=qqqqff] (23.5,14.5)-- (23.5,15.);
\draw [->] (20.,13.5) -- (20.5,14.);
\draw [->] (21.5,13.5) -- (21.,14.);
\draw [->] (22.274717458050993,14.238788650061045) -- (22.85863288478855,14.238788650061045);
\draw (19.6,15.42) node[anchor=north west] {\footnotesize{$1$}};
\draw (23.12,15.42) node[anchor=north west] {\footnotesize{$1$}};
\draw (20.9,14.95) node[anchor=north west] {\footnotesize{$14$}};
\draw (24.4,14.95) node[anchor=north west] {\footnotesize{$14$}};
\draw (20.55,15.6) node[anchor=north west] {\footnotesize{$s$}};
\draw (24.03,15.7) node[anchor=north west] {\footnotesize{$s'$}};
\draw (19.85,13.1) node[anchor=north west] {\footnotesize{$E(s)=-4$}};
\draw (23.35,13.1) node[anchor=north west] {\footnotesize{$E(s')=-7$}};
\begin{scriptsize}
\draw [fill=ffqqqq] (21.,14.5) circle (3.0pt);
\draw [fill=ffqqqq] (20.5,14.5) circle (3.0pt);
\draw [fill=ffqqqq] (20.5,15.) circle (3.0pt);
\draw [fill=ffqqqq] (21.,15.) circle (3.0pt);
\draw [fill=qqffqq] (21.5,15.) circle (3.0pt);
\draw [fill=ffqqqq] (21.5,14.5) circle (3.0pt);
\draw [fill=qqffqq] (21.5,14.) circle (3.0pt);
\draw [fill=ffqqqq] (21.5,13.5) circle (3.0pt);
\draw [fill=qqffqq] (21.,13.5) circle (3.0pt);
\draw [fill=qqffqq] (20.5,13.5) circle (3.0pt);
\draw [fill=ffqqqq] (20.,13.5) circle (3.0pt);
\draw [fill=qqffqq] (20.,14.) circle (3.0pt);
\draw [fill=ffqqqq] (20.,14.5) circle (3.0pt);
\draw [fill=ffqqqq] (20.,15.) circle (3.0pt);
\draw [fill=ffqqqq] (24.5,14.5) circle (3.0pt);
\draw [fill=ffqqqq] (24.,14.5) circle (3.0pt);
\draw [fill=ffqqqq] (24.,15.) circle (3.0pt);
\draw [fill=ffqqqq] (24.5,15.) circle (3.0pt);
\draw [fill=qqffqq] (25.,15.) circle (3.0pt);
\draw [fill=ffqqqq] (25.00044234327216,14.487543914502325) circle (3.0pt);
\draw [fill=qqffqq] (25.,14.) circle (3.0pt);
\draw [fill=ffqqqq] (24.50044234327216,13.987543914502325) circle (3.0pt);
\draw [fill=qqffqq] (24.5,13.5) circle (3.0pt);
\draw [fill=qqffqq] (24.,13.5) circle (3.0pt);
\draw [fill=ffqqqq] (24.,14.) circle (3.0pt);
\draw [fill=qqffqq] (23.5,14.) circle (3.0pt);
\draw [fill=ffqqqq] (23.5,14.5) circle (3.0pt);
\draw [fill=ffqqqq] (23.5,15.) circle (3.0pt);
\end{scriptsize}
\end{tikzpicture}
  \caption{An illustrative  example of the diagonal pull-move neighborhood in 2D square lattice model. }
\label{figure4}
\end{figure}
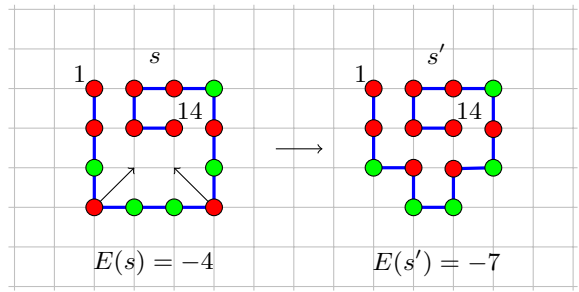
\end{center}
\begin{algorithm}[htp]
\setstretch{1}
\caption{The proposed Hill Climbing Algorithm: HCA}
\begin{flushleft}
\algorithmicrequire  A feasible solution  $s=(s_{1},\ldots,s_{n})$.\vspace{0.1cm} \\
\algorithmicensure  The best found  neighbor for $s$.\\ \vspace{-0.29cm}
\rule{\linewidth}{0.02pt}
\end{flushleft}
\begin{algorithmic}[htp]
\State \textbf{Begin}
 \State  Calculate $f(s);$
 \State $s^{*}=s;$
 \State  $iter=0$
 \While {$iter\leq$ Max (maximum number of iterations)}
\For {$i = 1,\ldots,n$}
 \State  $s'$= diagonal pull-move $(s,i);$
 \If {$f(s')< f(s)$ or $rand \leq \beta$}
 \State $s=s';$
 \EndIf \State \textbf{end if}
 \EndFor
 \State \textbf{end for}
 \If {$ f(s) < f(s^{*})$}
  \State $s^{*}=s;$
 \EndIf \State \textbf{end if}
\State $iter=iter+1;$
\EndWhile
\State \textbf{end while}
\State  \textbf{End}
\end{algorithmic}
\label{Algorithm3}
\end{algorithm}
\subsection{Rotation and Opposite motion}
Rotation and opposite motion are two local operations that are used in \cite{lin_using_2012}, to create new solutions by changing the movement directions of a given structure:
\begin{enumerate}
                    \item Opposite  motion: This motion allows to generate a new structure by simply reversing the directions of a given structure between two randomly chosen positions $i$ and $j$.
                    \item Rotational motion: the protein structure can rotate clockwise (CW) or counterclockwise (CCW). Similar to reverse motion, to generate a new solution from a given structure $s$, we change a substructure $s'$ (i.e., a local structure) in $s$ by rotating $s'$ that is limited between two randomly chosen positions $i$ and $j$ in  CW or CCW direction.
                  \end{enumerate}
\subsection{Step Length}
Here, the step length represents the length of the subsequence that will be rotated or inversed for a given conformation. The rotation rules used are described in \cite{lin_using_2012}.
\subsection{Generation of a New Solution Using  L\'{e}vy Flights}
One of the most important advantages of the random walk via L\'{e}vy Flights is its ability to explore the search space through its step length characteristics. In order to use this strategy for PFP, we associate the step length (i.e., the length of the subsequence) with the value generated by the L\'{e}vy Flights. For example, if $s$ is the protein sequence of $n$ amino acids, we divide the interval $[0,1]$ into $k$ subintervals and define the step length $step_{L}$ as follows:
\begin{table}[H]
\centering
\setlength{\tabcolsep}{1em}
\setlength{\arrayrulewidth}{.05em}
{\renewcommand{\arraystretch}{1.8}
\begin{tabular}{|l|c|l|l|l|}
\hline
\multicolumn{1}{|c|}{$u$} &  $[0,\frac{1}{k}[$ & \multicolumn{1}{c|}{$[\frac{1}{k}, \frac{2}{k}[$} & \ldots& \multicolumn{1}{c|}{[$\frac{k-1}{k}$, 1]} \\
\hline
\multicolumn{1}{|c|}{$step_{L} $} & [1,$\frac{L_{max}}{k}$[ & \multicolumn{1}{c|}{[$\frac{L_{max}}{k}$,  $\frac{2L_{max}}{k}$[} &\ldots & \multicolumn{1}{c|}{[$\frac{(k-1)L_{max}}{k}$ ,  $ L_{max}$]} \\
\hline
\end{tabular}}
\end{table}
\noindent where $L_{max}= \dfrac{n}{h}$, $h \in\{1,2,\ldots,n\}$, and $u$ is the value generated by L\'{e}vy Flights. If, for instance, $n=60$, $h=3$ and $k=4$, then $L_{max}=20$. So, we select randomly the length of the subsequence in the generated interval by L\'{e}vy Flights value according the following Table:
\begin{table}[htp]
\centering
\setlength{\tabcolsep}{1em}
\setlength{\arrayrulewidth}{.05em}
{\renewcommand{\arraystretch}{1.8}
\begin{tabular}{|l|c|l|l|l|}
\hline
\multicolumn{1}{|c|}{$u$} &  [0, 0.25[ & \multicolumn{1}{c|}{ [0.25, 0.5[} & \multicolumn{1}{c|}{[0.5, 0.75[} & \multicolumn{1}{c|}{[0.75, 1]} \\
\hline
\multicolumn{1}{|c|}{$step_{L} $} & [1,5[ & \multicolumn{1}{c|}{[5, 10[} & \multicolumn{1}{c|}{[10, 15[} & \multicolumn{1}{c|}{[15, 20]} \\
\hline
\end{tabular}}
\end{table}

Let $i$ be a random value between $1$ and $n-step_{L}$. The rotated subsequence $sb$ is then defined as: \mbox{$sb=s_{i}s_{i+1}\ldots s_{i+step_{L}}$}.
\subsection{Rotation of represented solutions }
This operation does not change the energy value of the solution because all confirmation will be rotated. It allows increasing the diversity of the population by rotating a part of the represented solutions. This technique will enhance the chances for the crossover operator to generate feasible solutions in the next generation.
\section{Experimental Results and Discussion}
 The objective of this section is to demonstrate the effectiveness of the proposed HCSA algorithm and to compare its performance with a set of well-known effective approaches that have been published in the literature for the HP protein folding problem in the 3D cubic lattice model. Our suggested algorithm has been tested on two benchmark instance datasets. Table \ref{table1} contains a set of standard 3D HP benchmark instances with different sizes that have been widely used in the literature for solving the PFP in the 3D cubic lattice model, which was introduced in \cite{jiang_protein_2003, garza-fabre_multi-objectivization_2015, garza-fabre_comparative_2013}. However, the instances reported in the Table \ref{table2} are a set of short protein sequences that contain 27 amino acids \cite{guo_simulation_2006}. The last column of both tables gives the Best Known energy Value (BKV) for the corresponding instance in the 3D cubic lattice. For each instance, we performed 30 independent runs. The parameters used in the HCSA method are fixed as follows:
  \begin{enumerate}
    \item The population size $N=50$ for sequences of length $n\leq 36$, $N=100$ for $36 <n\leq 50$ and $N=150$ for $n > 50$.\vspace{0.2em}
    \item The number of generations  is 500.\vspace{0.2em}
    \item A cuckoo changes its direction with probability (rotation probability) $p_{c}=0.3$.\vspace{0.2em}
    \item A cuckoo egg is discovered by the host bird with probability $p_{a}=0.25$.\vspace{0.2em}
    \item The value of the parameter $\beta$ for the HCA is $\beta=0.2$.\vspace{0.2em}
    \item  For the cuckoo's step length, the parameters are fixed at $\lambda = \frac{3}{2}$, $\alpha = 1$.\vspace{0.2em}
  \end{enumerate}

\begin{table}[H]
\centering
\setlength{\tabcolsep}{1em}
\setlength{\arrayrulewidth}{.05em}
{\renewcommand{\arraystretch}{1.2}
\begin{footnotesize}
\begin{tabular}{|l|l|l|l|}
\hline
\multicolumn{1}{|c|}{Seq.} & \multicolumn{1}{c|}{Length} & \multicolumn{1}{c|}{Sequence} & BKV \\
\hline\hline
\multicolumn{1}{|c|}{$s_{1}$} & \multicolumn{1}{c|}{20} & \multicolumn{1}{c|}{$(HP)^{2}PH(HP)^{2}(PH)^{2}HP(PH)^{2}$} & -11 \\
\hline
\multicolumn{1}{|c|}{$s_{2}$} & \multicolumn{1}{c|}{24} & \multicolumn{1}{c|}{$H^{2}P^{2}(HP^{2})^{6}H^{2}$} & -13 \\
\hline
\multicolumn{1}{|c|}{$s_{3}$} & \multicolumn{1}{c|}{25} & \multicolumn{1}{c|}{$P^{2}HP^{2}(H^{2}P^{4})^{3}H^{2}$} & -9 \\
\hline
\multicolumn{1}{|c|}{$s_{4}$} & \multicolumn{1}{c|}{36} & \multicolumn{1}{c|}{$P(P^{2}H^{2})^{2}P^{5}H^{5}(H^{2}P^{2})^{2}P^{2}H(HP^{2})^{2}$} & -18 \\
\hline
\multicolumn{1}{|c|}{$s_{5}$} & \multicolumn{1}{c|}{46} & \multicolumn{1}{c|}{$P^{2}H^{3}PH^{3}P^{3}HPH^{2}PH^{2}P^{2}HPH^{4}PHP^{2}H^{5}PHPH^{2}P^{2}H^{2}P$} & -35 \\
\hline
\multicolumn{1}{|c|}{$s_{6}$} & \multicolumn{1}{c|}{48} & \multicolumn{1}{c|}{$P^{2}H(P^{2}H^{2})^{2}P^{5}H^{10}P^{6}(H^{2}P^{2})^{2}HP^{2}H^{5}$} & -31 \\
\hline
\multicolumn{1}{|c|}{$s_{7}$} & \multicolumn{1}{c|}{50} & \multicolumn{1}{c|}{ $H^{2}(PH)^{3}PH^{4}PH(P^{3}H)^{2}P^{4}(HP^{3})^{2}HPH^{4}(PH)^{3}PH^{2}$} & -34 \\
\hline
\multicolumn{1}{|c|}{$s_{8}$} & \multicolumn{1}{c|}{58} & \multicolumn{1}{c|}{$PH(PH^{3})^{2}P(PH^{2}PH)^{2}H(HP)^{3}(H^{2}P^{2}H)^{2}PHP^{4}(H(P^{2}H)^{2})^{2}$} & -44 \\
\hline
\multicolumn{1}{|c|}{$s_{9}$} & \multicolumn{1}{c|}{60} & \multicolumn{1}{c|}{ $P(PH^{3})^{2}H^{5}P^{3}H^{10}PHP^{3}H^{12}P^{4}H^{6}PH^{2}PHP$} & -55 \\
\hline
\multicolumn{1}{|c|}{$s_{10}$} & \multicolumn{1}{c|}{64} & \multicolumn{1}{c|}{$H^{12}(PH)^{2}((P^{2}H^{2})^{2}P^{2}H)^{3}(PH)^{2}H^{11}$} & -59 \\
\hline
\multicolumn{1}{|c|}{$s_{11}$} & \multicolumn{1}{c|}{67} & \multicolumn{1}{c|}{$P(HPH^{2}PH^{2}PHP^{2}H^{3}P^{3})^{3}(HPH)^{3}P^{2}H^{3}P$} & -56 \\
\hline
\multicolumn{1}{|c|}{$s_{12}$} & \multicolumn{1}{c|}{88} & \multicolumn{1}{c|}{$P(HPH)^{3}P^{2}H^{2}(P^{2}H)^{6}H(P^{2}H^{3})^{4}P^{2} (HPH)^{3} P^{2}HP(PHP^{2}H^{2}P^{2}HP)^{2}$} & -72 \\
\hline
\multicolumn{1}{|c|}{$s_{13}$} & \multicolumn{1}{c|}{103} & \multicolumn{1}{c|}{$P^{2}H^{2}P^{5}H^{2}P^{2}H^{2}PHP^{2}HP^{7}HP^{3}H^{2}PH^{2}P^{6}HP^{2}HPHP^{2}HP^{5}H^{3}P^{4}$} & -85 \\
\multicolumn{1}{|c|}{} & \multicolumn{1}{c|}{} & \multicolumn{1}{c|}{$H^{2}PH^{2}P^{5}H^{2}P^{4}H^{4}PHP^{8}H^{5}P^{2}HP^{2}$} &  \\
\hline
\multicolumn{1}{|c|}{$s_{14*}$} & \multicolumn{1}{c|}{124} & \multicolumn{1}{c|}{$P^{3}H^{3}PHP^{4}HP^{5}H^{2}P^{4}H^{2}P^{2}H^{2}(P^{4}H)^{2} P^{2}HP^{2}H^{2}P^{3}H^{2}PHPH^{3}$} & -75 \\
\multicolumn{1}{|c|}{} & \multicolumn{1}{c|}{} & \multicolumn{1}{c|}{$P^{4}H^{3}P^{6}H^{2}P^{2} HP^{2}HPHP^{2}HP^{7}HP^{2}H^{3}P^{4}HP^{3}H^{5}P^{4}H^{2}(PH)^{4}$} &  \\
\hline
\multicolumn{1}{|c|}{$s_{15}$} & \multicolumn{1}{c|}{136} & \multicolumn{1}{c|}{$HP^{5}HP^{4}HPH^{2}PH^{2}P^{4}HPH^{3}P^{4}HPHPH^{4}P^{11} HP^{2}HP^{3}HPH^{2}P^{3}H^{2}P^{2}$} & -83 \\
\multicolumn{1}{|c|}{} & \multicolumn{1}{c|}{} & \multicolumn{1}{c|}{$HP^{2}HPHPHP^{8}HP^{3} H^{6}P^{3}H^{2}P^{2}H^{3}P^{3}H^{2}PH^{5}P^{9}HP^{4}HPHP^{4}$} &  \\
\hline
\end{tabular}
\caption{The 2D and 3D HP standard  benchmark instances.}
\label{table1}
\end{footnotesize}}
\end{table}
\begin{table}[H]
\centering
\setlength{\tabcolsep}{1.5em}
\setlength{\arrayrulewidth}{.05em}
{\renewcommand{\arraystretch}{1.2}
\begin{footnotesize}
\begin{tabular}{|c|c|c|c|}
\hline
Seq. & Length  & Protein sequence in the H-P model &  BKV\\
\hline\hline
$A_{1}$ & 27 & $PHPHPH^{3}P^{2}HPHP^{11}H^{2}P$ & -9 \\
\hline
$A_{2}$ & 27 & $PH^{2}P^{10}H^{2}P^{2}H^{2}P^{2}HP^{2}HPH$ & -10 \\
\hline
$A_{3}$ & 27 & $ H^{4}P^{5}HP^{4}H^{3}P^{9}H $& -8 \\
\hline
$A_{4}$ & 27 & $ H^{3}P^{2}H^{4}P^{3}HPHP^{2}H^{2}P^{2}HP^{3}H^{2} $& -15 \\
\hline
$A_{5}$ & 27 & $ H^{4}P^{4}HPH^{2}P^{3}H^{2}P^{10}$ & -8 \\
\hline
$A_{6}$ & 27 & $ HP^{6}HPH^{3}P^{2}H^{2}P^{3}HP^{4}HPH $& -12 \\
\hline
$A_{7}$ &  27 & $ HP^{2}HPH^{2}P^{3}HP^{5}HPH^{2}PHPHPH^{2} $& -13 \\
\hline
$A_{8}$ &  27 & $ HP^{11}HPHP^{8}HPH^{2} $& -4 \\
\hline
$A_{9}$ &  27 & $ P^{7}H^{3}P^{3}HPH^{2}P^{3}HP^{2}HP^{3} $& -7 \\
\hline
$A_{10}$ &  27 & $ P^{5}H^{2}PHPHPHPHP^{2}H^{2}PH^{2}PHP^{3} $& -11 \\
\hline
$A_{11}$ &  27 & $ HP^{4}H^{4}P^{2}HPHPH^{3}PHP^{2}H^{2}P^{2}H  $& -16 \\
\hline
\end{tabular}
\vspace{0.05cm}
\caption{Instances  with 27 amino acids used in our experiment for 3D cubic lattice model.}
\label{table2}
\end{footnotesize}}
\end{table}
\subsection{Comparison of the best prediction}
In Table \ref{table3}, we perform a comparison between the lowest free energy obtained for the 3D H-P reference instances in Table \ref{table1}, by the proposed algorithm and those achieved by the Genetic Algorithm (GA column) \cite{garza-fabre_locality-based_2012-1}, the Evolutionary Algorithm (EA column) \cite{garza-fabre_multi-objectivization_2015}. The Ant Colony Optimization algorithm (ACO column) \cite{thilagavathi_aco-metaheuristic_2015}, a hybrid approach that combines the Genetic Algorithm with the Particle Swarm Optimization algorithm (HGA-PSO column) \cite{lin_protein_2011},  two variants of the Immune Algorithm (AIS \cite{cutello_immune_2005}, CI \cite{de_almeida_hybrid_2007}), the Best Improvement Local Search (BILS) algorithm \cite{garza-fabre_comparative_2013}, and the Heuristic Algorithm (EHA) that was introduced in \cite{traykov_algorithm_2018}. According to the results given in Table 3, it is obvious and clear that HCSA is able to obtain the optimal conformation for sequences with length less than 50 (i.e., instances ranked from $s_{1}$ to $s_{6}$), and a near-optimal solution for the rest of the instances. Moreover, there is no significant difference in terms of the best energy value that was obtained by our HCSA and the BKV for most of the instances. For example, the best energy value obtained by the HCSA for the sequence $s_{6}$ that contains 60 amino acids is -54, while the BKV for the same sequence is -55. As we can observe in Table \ref{table3}, the majority of the compared algorithms are able to achieve the optimal conformation for short-length instances (instances containing less than 36 amino acids).
\begin{table}[htp]
\centering
\setlength{\tabcolsep}{0.62em}
\setlength{\arrayrulewidth}{.07em}
{\renewcommand{\arraystretch}{1.7}
\begin{footnotesize}
\begin{tabular}{|cccccccccccc|}
\hline
Seq. &   Length &  BKV & ACO & EHA & HGA-PSO & AIS & CI & EA & GA &                  BILS  &  HCSA \\
\hline\hline
$3d1$ &  20 & \textbf{-11} & -10 & \textbf{-11} & \textbf{-11} & \textbf{-11} & \textbf{-11} & \textbf{-11} & \textbf{-11} & -10  &  \textbf{-11} \\
\hline
$3d2$ &     24 &  \textbf{-13} & -8 & \textbf{-13} & \textbf{-13} & \textbf{-13} & \textbf{-13} & \textbf{-13} & \textbf{-13} &   -9  &\textbf{-13} \\
\hline
$3d3$ &     25 & \textbf{-9} &-6 & \textbf{-9} & \textbf{-9} & \textbf{-9} & \textbf{-9} & \textbf{-9} & \textbf{-9} &  -7 & \textbf{-9} \\
\hline
$3d4$ &    36 & \textbf{-18} & -10 & \textbf{-18} & \textbf{-18} & \textbf{-18} & \textbf{-18} & \textbf{-18} & \textbf{-18} &  -12  &\textbf{-18} \\
\hline
$3d5$ &     46 & \textbf{-35} & -21 & -32 & NA & NA & NA & -30 & -32 & -22  &   \textbf{-33} \\
\hline
$3d6$ &    48 & \textbf{-31} & NA & \textbf{-31} & -29 & -29 & -29 & -29 & -31 & -19 & \textbf{-31} \\
\hline
$3d7$ &    50 & \textbf{-34} & NA & -28 & -26 & -23 & -26 & -25 & -30 & -18 & \textbf{-31} \\
\hline
$3d8$ &    58 & \textbf{-44} & NA & NA & NA & NA & NA & -35 & -37 & -23 & \textbf{-41} \\
\hline
$3d9$ &    60 & \textbf{-55} & NA & \textbf{-55} & -49 & -41 & -48 & -48 & -50 &  -36  &\textbf{-54} \\
\hline
$3d10$ &   64 & \textbf{-59} & NA & NA & NA & -42 & NA & -45 & -52 & -34 & \textbf{-57} \\
\hline
$3d11$ &   67 & \textbf{-56} & NA & NA & NA & NA & NA & -40 & -41 &-28 & \textbf{-45} \\
\hline
$3d12$ &   88 & \textbf{-72} & NA & NA & NA & NA & NA & -48 & -50 & -29 & \textbf{-57} \\
\hline
$3d13$ &   103 &  \textbf{-85} & NA & NA & NA & NA & NA & -41 & -41 &-22 & \textbf{-47} \\
\hline
$3d14$ &     124 & \textbf{-75} & NA & NA & NA & NA & NA & -48 & -51 & -24 & \textbf{-57} \\
\hline
$3d15$ &     136 & \textbf{-83} & NA & NA & NA & NA & NA & -51 & -52 & -30 & \textbf{-65} \\
\hline
 \multicolumn{10}{l}{Bold values are the best energy for the corresponding instance.}\\
  \multicolumn{10}{l}{NA is referring to unavailable data in the literature.}
\end{tabular}
\caption{The best results obtained by HCSA compared with state-of-the-art approaches for 15 H-P sequences in 3D cubic lattice.}
\label{table3}
\end{footnotesize}}
\end{table}\\
In comparison, the suggested algorithm outperforms better than the other mentioned approaches with a considerable difference for instances containing more than 50 amino acids (i.e., for instances ranging from $s_{5}$ to $s_{15}$), in particular, when compared to the GA, EA, AIS and IBLS algorithms. On the other hand, we can also see that EHA and EA are the two most competitive algorithms against HCSA.\\
 To properly evaluate our HCSA algorithm performance, we also tested it using the instances of the HP 3D benchmark presented in Table \ref{table2}. We have compared the best results obtained by HCSA with those of three metaheuristic algorithms known in the literature, namely the Genetic Algorithm (GA) \cite{custodio_investigation_2004}, two-hybrid Particle Swarm Optimization methods ($TPPSO^{1}$, $TPPSO^{2}$) \cite{guo_hybrid_2017}, and the Elastic Network Algorithm (EN) \cite{guo_simulation_2006}. Based on the results in Table \ref{table4}, we can see that the suggested algorithm achieved the optimal energy value for the 11 benchmark instances. In comparison, these results show that HCSA and $TPPSO^{2}$ algorithms have better performance than EN, GA, $TPPSO^{1}$, only HCSA and $TPPSO^{2}$, can get the optimal free energy of each instance. However, EN, $GA$, and $TPPSO^{1}$ can not obtain the lowest free energy of some instances, in particular, for $A_{1}$, $A_{2}$, $A_{6}$, $A_{7}$ and $A_{11}$. Figure 10 shows the best 3D conformations obtained by the suggested algorithm for the instances $A_{6}$, $A_{7}$, and $A_{11}$.
\begin{table}[H]
\centering
\setlength{\tabcolsep}{1em}
\setlength{\arrayrulewidth}{.07em}
{\renewcommand{\arraystretch}{1.2}
\begin{footnotesize}

\begin{tabular}{|cccccccc|}
\hline
\textbf{Seq.} & \textbf{Length} & \textbf{BKV} & \textbf{EN} & \textbf{GA} & \textbf{$TPPSO^{1}$} & \textbf{$TPPSO^{2}$} & \textbf{HCSA} \\
\hline
\hline
$A_{1}$ & 27 & \textbf{-9} & \textbf{-9} & -8 & \textbf{-9} & \textbf{-9} & \textbf{-9} \\
\hline
$A_{2}$ & 27 & \textbf{-10} & \textbf{-10} & \textbf{-10} & \textbf{-10} & \textbf{-10} & \textbf{-10} \\
\hline
$A_{3}$ & 27 & \textbf{-8} & \textbf{-8} & \textbf{NA} & \textbf{-8} & \textbf{-8} & \textbf{-8} \\
\hline
$A_{4}$ & 27 & \textbf{-15} & \textbf{-15} & \textbf{-15} & \textbf{-15} & \textbf{-15} & \textbf{-15} \\
\hline
$A_{5}$ & 27 & \textbf{-8} & \textbf{-8} & \textbf{-8} & \textbf{-8} & \textbf{-8} & \textbf{-8} \\
\hline
$A_{6}$ & 27 & \textbf{-12} & \textbf{-12} & \textbf{NA} & -11 & \textbf{-12} & \textbf{-12} \\
\hline
$A_{7}$ & 27 & \textbf{-13} & \textbf{-13} & \textbf{-13} & -12 & \textbf{-13} & \textbf{-13} \\
\hline
$A_{8}$ & 27 & \textbf{-4} & \textbf{-4} & \textbf{-4} & \textbf{-4} & \textbf{-4} & \textbf{-4} \\
\hline
$A_{9}$ & 27 & \textbf{-7} & \textbf{-7} & \textbf{-7} & \textbf{-7} & \textbf{-7} & \textbf{-7} \\
\hline
$A_{10}$ & 27 & \textbf{-11} & \textbf{-11} & \textbf{NA} & \textbf{-11} & \textbf{-11} & \textbf{-11} \\
\hline
$A_{11}$ & 27 & \textbf{-16} & -14 & \textbf{NA} & -14 & \textbf{-16} & \textbf{-16} \\
\hline
\multicolumn{8}{l}{Bold values are the best energy for the corresponding instance.}\\
  \multicolumn{8}{l}{NA is referring to unavailable data in the literature.}
\end{tabular}
\caption{The best results obtained by HCSA compared with state-of-the-art approaches for 11 H-P sequences given in Table \ref{table2}.}
\label{table4}
\end{footnotesize}}
\end{table}
\begin{figure}[H]
   \includegraphics[width=5.5cm, height=5.7cm]{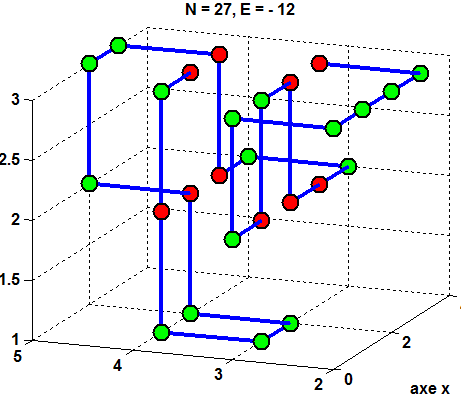}\includegraphics[width=5.5cm, height=5.7cm]{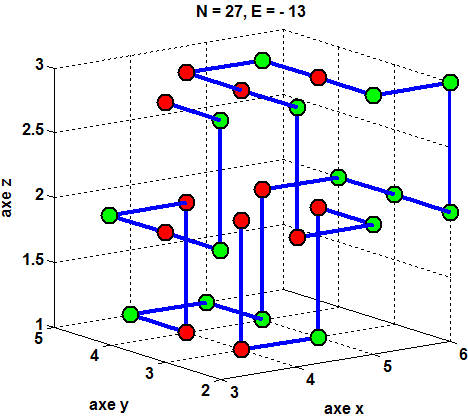}\includegraphics[width=5.5cm, height=5.7cm]{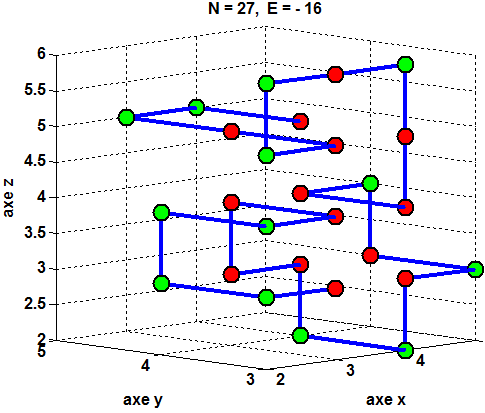}\\
  \caption{The best conformations obtained by HCSA  in 3D cubic lattice model for the instances $A_{6}$, $A_{7}$ and $A_{6}$.  }\label{figure10}
\end{figure}
\subsection{Comparison of Stability}
To demonstrate the stability of our algorithm for the 3D-HP lattice model, a comparison of the statistical results obtained with the EA and IA algorithms and our algorithm after the 50 independent runs for each instance considered is reported in Table \ref{table5}. In these experiments, we use the best and average energy values (ave. column) as metric values.
\begin{table}[H]
\centering
\setlength{\tabcolsep}{0.9em}
\setlength{\arrayrulewidth}{0.07em}
{\renewcommand{\arraystretch}{1.2}
\begin{footnotesize}
\begin{tabular}{|c|c|c|c|c|c|c|c|c|c|c|}
\hline
 & \multicolumn{2}{c|}{IA} & \multicolumn{2}{c|}{BILS} & \multicolumn{2}{c|}{EA} & \multicolumn{2}{c|}{GA} & \multicolumn{2}{c|}{HCSA} \\
\hline
Seq. & Best & Mean & Best & Mean & Best & Mean & Best & Mean & Best & Mean \\
\hline
3d1 & \textbf{-11} & -10.32 & -10 & -5.8 & \textbf{-11} & -10.6       & \textbf{-11} & \textbf{-11}  & \textbf{-11}  & \textbf{-11} \\
\hline
3d2 & \textbf{-13} & -10.90 & -9 & -5.2 & \textbf{-13} & -11.6         & \textbf{-13} & -12.94 & \textbf{-13} & \textbf{-13} \\
\hline
3d3 & \textbf{-9} & -7.98 & -7 & -2.7 & \textbf{-9} & -8.7              & \textbf{-9} & -8.71 & \textbf{-9} &  \textbf{-9}\\
\hline
3d4 & \textbf{-18} & -14.38 & -12 & -6.5 & \textbf{-18} & -15.4         & \textbf{-18} & -15.91 & \textbf{-18} & \textbf{-18} \\
\hline
3d5 & NA & NA & -22 & -13.9 & -30 & -24.4                              & -32 & -27.72 & \textbf{-33} & \textbf{-30.25}  \\
\hline
3d6 & -29 & -20.80 & -19 & -12.4 & -29 & -23.2                         & \textbf{-31} & -26.59 & \textbf{-31} & \textbf{-28.90} \\
\hline
3d7 & -23 & -20.20 & -18 & -11.8 & -25 & -21.1                          & -30 & -26.43 & \textbf{-31} &  \textbf{-28.95}\\
\hline
3d8 & NA & NA & -29 & -15.9 & -35 & -27.7                               & -37 & -32.39 & \textbf{-41} &  \textbf{-36.20}\\
\hline
3d9 & -41 & -34.18 & -36 & -25.8 & -48 & -38.1                          & -50 & -43.46 & \textbf{-54} & \textbf{-48.25} \\
\hline
3d10 & -42 & -33.01 & -34 & -24.8 & -45 & -36.2                           & -52 & -46.12 & \textbf{-57} & \textbf{-50.20} \\
\hline
3d11 & NA & NA & -28 & -18.2 & -40 & -31.0                               & -41 & -36.39 & \textbf{-45} & \textbf{-40.65} \\
\hline
3d12 & NA & NA & -29 & -20.6 & -48 & -37.3                               & -50 & -44.02 & \textbf{-57} &  \textbf{-50.95}\\
\hline
3d13 & NA & NA & -22 & -13.0 & -41 & -30.7                                & -41 & -34.99 & \textbf{-47} &  \textbf{-39.45} \\
\hline
3d14 & NA & NA & -24 & -16.6 & -48 & -35.5                               & -51 & -41.83 & \textbf{-57} & \textbf{-49.40} \\
\hline
3d15 & NA & NA & -30 & -19.2 & -51 & -38.6                               & -52 & -45.51 & \textbf{-65} & \textbf{-52.57} \\
\hline
\multicolumn{10}{l}{Bold values are the best energy for the corresponding instance.}\\
  \multicolumn{10}{l}{NA is referring to unavailable data in the literature.}
\end{tabular}
\end{footnotesize}}
\caption{Experimental results regarding the convergence and stability of HCSA compared against a set of state-of-the-art approaches, namely GA, EA, IA and BILS algorithms, for standard HP benchmark instances in a 3D cubic lattice.}\label{table5}
\end{table}
Table \ref{table5},  shows clearly that the best and average energy values obtained by the suggested algorithm HCSA are lower with a significant difference than those achieved by EA, GA, IBLS, and IA algorithms for all instances. Furthermore, the proposed algorithm achieved average energy values lower than the best one obtained by the other mentioned algorithms for the majority of instances. Moreover,  the suggested algorithm achieved a convergence rate equal to $100\%$ for 6 instances over 15 rangings from $s_{1}$ to $s_{6}$. The results demonstrate that the proposed HCSA is more efficient than state-of-art algorithms in terms of the quality of solutions for the protein folding problem in the 3D-HP cubic lattice model. Similarly, these results demonstrate the stability of our algorithm.

\subsection{Effect of the Hill-Climbing  Algorithm and l\'{e}vy Flight Step}
In order to show the effect of the improvement phase and the L\'{e}vy flights step that we used in the proposed algorithm. We have implemented the proposed algorithm with and without the proposed Hill-Climbing  Algorithm (HCA), the L\'{e}vy Flights Step (LFS), where we compare the results obtained by both versions with the suggested HCSA, we used the same initial population that contains 200 random solutions. We have also used the same values for the l\'{e}vy flights parameters.  The results shown in Table 6 are achieved through 50 independent runs for each instance.
\begin{table}[H]
\centering
\setlength{\tabcolsep}{1em}
\setlength{\arrayrulewidth}{0.08em}
{\renewcommand{\arraystretch}{1.6}
\begin{footnotesize}

\begin{tabular}{|c|clc|clc|clc|}
\cline{2-10}
\multicolumn{1}{c|}{} & \multicolumn{3}{c|}{HCSA} & \multicolumn{3}{c|}{HCSA without HCA} & \multicolumn{3}{c|}{HCSA without LFS } \\
\hline
Seq. & Best &  & Mean & Best &  & Mean & Best &  & Mean \\
\cline{1-1} \cline{2-2}\cline{4-5}\cline{7-7}\cline{8-8}\cline{10-10}
$A_{1}$ & \textbf{-9} &  & \textbf{-9} & -8 &  & -7.11 & -8 &  & -7.32 \\
$A_{2}$ & \textbf{-10} &  & \textbf{-10} & -9 &  & -8.72 & \textbf{-10} &  & -7.94 \\
$A_{3}$ & \textbf{-8} &  & \textbf{-7.89} & \textbf{-8} &  & -6.44 & \textbf{-8} &  & -6.62 \\
$A_{4}$ & \textbf{-15} &  & \textbf{-14.67} & -14 &  & -12.84 & -12 &  & -10.28 \\
$A_{5}$ & \textbf{-8} &  & \textbf{-8} & -7 &  & -6.04 & -7 &  & -6.37 \\
$A_{6}$ & \textbf{-12} &  & \textbf{-11.22} & -11 &  & -9.43 & -10 &  & -7.54 \\
$A_{7}$ & \textbf{-13} &  & \textbf{-12.04} & -12 &  & -10.02 & -12 &  & -10.2 \\
$A_{8}$ & \textbf{-4} &  & \textbf{-4} & \textbf{-4} &  & \textbf{-4} & \textbf{-4} &  & -3.92 \\
$A_{9}$ & \textbf{-7} &  & \textbf{-7} & \textbf{-7} &  & -5.87 & \textbf{-7} &  & -6.04 \\
$A_{10}$ & \textbf{-11} &  & \textbf{-10.37} & \textbf{-11} &  & -9.33 & \textbf{-11} &  & -9.18 \\
$A_{11}$ & \textbf{-16} &  & \textbf{-14.98} & -14 &  & -11.58 & -14 &  & -11.07 \\
\hline
\end{tabular}
\label{table6}
\end{footnotesize}}
\vspace{0.5cm}
\caption{Experimental results concerning the effect of the Hill-Climbing Algorithme (HCA) and,  l\'{e}vy Flight Step (LFS) that we have used in our HCSA algorithm.}

\end{table}
The results presented in Table 6 clearly illustrate the effect of HCA and LFS strategy on the stability and convergence of the HCSA algorithm. Only the algorithm that combines the two techniques achieves the optimal solution for all tested instances. In addition, these results also show a strong difference between HCSA and the other two versions in terms of average solutions. The stability of HCSA is justified by the balance between diversification and intensification of the search process that is provided by the LFS and HCA.

%
%
%
%
\begin{figure}[H]
   \includegraphics[  width=7.7cm, height=5.5cm]{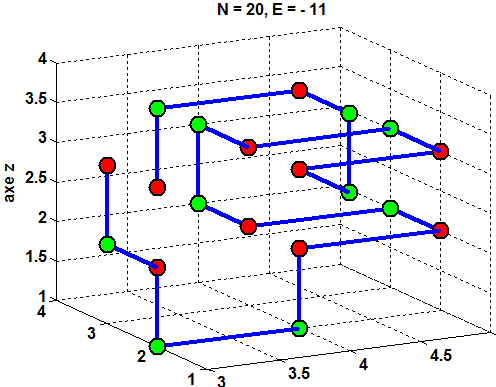}\includegraphics[width=7.7cm, height=5.5cm]{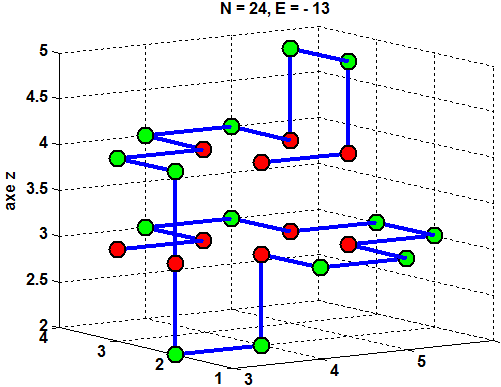}\\
  \includegraphics[width=7.7cm, height=6cm]{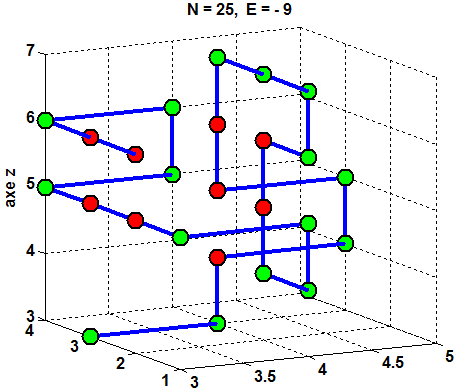}\includegraphics[width=7.7cm, height=6cm]{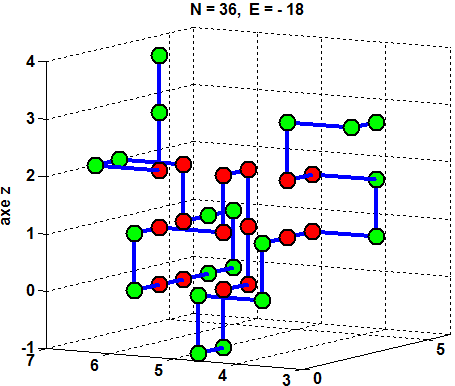}\\
  \includegraphics[width=7.7cm, height=6.3cm]{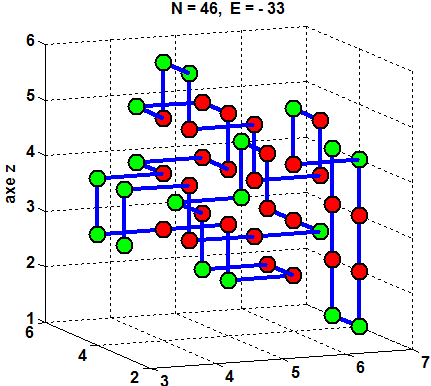}\includegraphics[width=7.7cm, height=6.3cm]{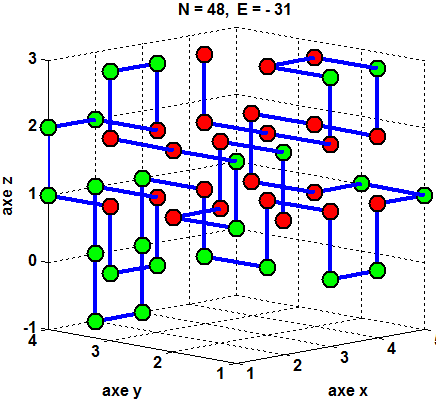}\\
\includegraphics[width=7.7cm, height=6cm]{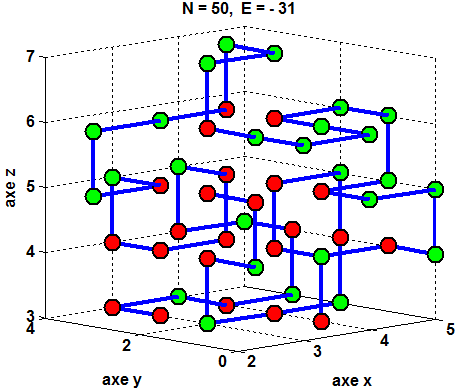}\includegraphics[width=7.7cm, height=6cm]{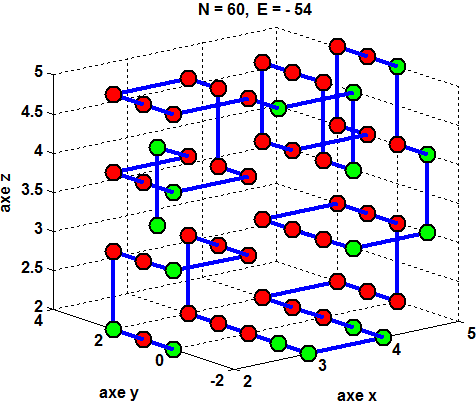}\\
  \caption{The best conformations obtained by HCSA for the HP benchmark instances in 3D cubic lattice model.  }\label{figure11}
\end{figure}
\section{Conclusion}
In this paper, we have presented an efficient Hybrid Cuckoo Search optimization Algorithm (HCSA), which we have applied to the H-P protein folding problem in a 3D lattice model. This hybridization combines the cuckoo search algorithm that using the L\'{e}vy Flights as an exploration strategy and the crossover operation and the Hill-Climbing Algorithm to improve the quality of the solutions. The simulation results on a set of data benchmark instances show that the suggested algorithm outperforms better than the state-of-the-art algorithms in the 3D lattice model in terms of the best results. similarly, we have compared the stability of our suggested algorithm against state-of-the-art algorithms using the best and average results as metrics values, this results indicated the superiority of HCSA to other test algorithms.  We believe that using of LFS improves the performance of the proposed algorithm, in particular, by its ability to avoid the stagnation of HCSA at a local optimum. As future work, it is very interesting to use the same algorithm to solve the PFP problem in other types of lattices such as the 2D and 3D triangular lattices. In addition, we can combine the cuckoo algorithm with other metaheuristics such as the Particle Swarm Optimization algorithm to solve other combinatorial optimization problems.

\section{Acknowledgement }

\end{document}